\documentclass{emulateapj}
\shorttitle{Aliases}
\shortauthors{Dawson \& Fabrycky}
\usepackage{natbib}

\begin{document}

\title{Radial velocity planets de-aliased.  A new, short period for Super-Earth 55~Cnc~e}

\author{
Rebekah I. Dawson\altaffilmark{1}
 \&
Daniel C. Fabrycky\altaffilmark{2}
}

\affil{Harvard-Smithsonian Center for Astrophysics \\ 
60 Garden St, MS-10, Cambridge, MA 02138}

\altaffiltext{1}{{\tt  rdawson@cfa.harvard.edu}}
\altaffiltext{2}{Michelson Fellow; {\tt  daniel.fabrycky@gmail.com}}

\begin{abstract}
Radial velocity measurements of stellar reflex motion have revealed many extrasolar planets, but gaps in the observations produce aliases, spurious frequencies that are frequently confused with the planets' orbital frequencies. In the case of Gl~581~d, the distinction between an alias and the true frequency was the distinction between a frozen, dead planet and a planet possibly hospitable to life \citep{2007U,2009M}. To improve the characterization of planetary systems, we describe how aliases originate and present a new approach for distinguishing between orbital frequencies and their aliases. Our approach harnesses features in the spectral window function to compare the amplitude and phase of predicted aliases with peaks present in the data. We apply it to confirm prior alias distinctions for the planets GJ~876~d and HD~75898~b. We find that the true periods of Gl~581~d and HD~73526~b/c remain ambiguous. We revise the periods of HD~156668~b and 55~Cnc~e, which were afflicted by daily aliases.  For HD~156668~b, the correct period is 1.2699 days and minimum mass is $(3.1 \pm 0.4)$~$M_\oplus$. For 55~Cnc~e, the correct period is 0.7365 days -- the shortest of any known planet -- and minimum mass is $(8.3 \pm 0.3)$~$M_{\oplus}$. This revision produces a significantly improved 5-planet Keplerian fit for 55 Cnc, and a self-consistent dynamical fit describes the data just as well.  As radial velocity techniques push to ever-smaller planets, often found in systems of multiple planets, distinguishing true periods from aliases will become increasingly important. 
\end{abstract}


\section{Introduction}
\label{sec:intro}

In the past two decades, over 400 extrasolar planets have been discovered, including more than 300 detected by radial velocity measurements. The entire architecture of a planetary system is encoded in the wobbles of its host star. In frequency space, the star's radial velocity variations are decomposed into the frequencies associated with each planet's gravitational interactions. One obstacle in correctly attributing these frequencies to planets are the spurious alias frequencies in the periodogram of the star's radial velocity measurements, caused by the discrete time sampling of the observations. Convolved with the orbital frequencies of alien worlds are Earth's own rotational and orbital frequencies, which dictate when the host star is visible at night, and -- for many data sets -- the synodic lunar frequency, which impacts the allocation of telescope time. 

Distinguishing aliases from physical frequencies is a common problem, yet making the correct distinction is crucial for characterizing extrasolar planets. For example, \citet{2007U} announced a super-Earth orbiting the M star Gl~581 with period 83 days, beyond the cold edge of the habitable zone. After more than doubling the number of observations, they determined that the planet's period was actually 67 days, well within the habitable zone, and that the 83 day period was an alias \citep{2009M}. The distinction between an alias and physical frequency was the distinction between a frozen, dead planet and a planet  possibly hospitable to life. For reasons we will describe below, planets with periods of one to several months -- in or near the habitable zone of M stars -- will typically have aliases with periods within about 30 days of their own orbital period. As more planets are discovered orbiting M stars, astronomers will be struggling to distinguish which of two close frequencies, one of which places the planet in the habitable zone, corresponds to a planet's orbital frequency. In general, planets with periods between a few months and a few years often have confusing aliases caused by convolution with Earth's orbital period, while planets with periods near a day, such as the super-Earth GJ~876~d \citep{2005R}, have confusing aliases caused by convolution with Earth's rotational period. Automatic de-aliasing algorithms, such as CLEAN \citep{1987R}, have been applied to particularly complicated radial velocity periodograms with some success \citep{2009Q}, yet, while they are good for cleaning up a periodogram, they should not be relied on for distinguishing between an alias and a physical frequency.  Aliases also pose a challenge for observing variable stars and period-searching algorithms have been designed to not fall prey to them (see for example \citealt{2008P, 2007Reegen,2010R}).

Therefore, to enhance detection and characterization of planets, we have developed an approach to identify aliases by harnessing features of the ``spectral window function," the Fourier transform of the observation times. Consider the star's motion as a signal that passes through a system, the time sampling window. Because of noise and loss of information, we can never perfectly reconstruct the signal. But we know everything there is to know about the system: for a sinusoid of a given amplitude, frequency, and phase, peaks in the window function cause aliases with calculable amplitudes and phases \citep{1975D,1976D}. The several time sampling frequencies -- sidereal year, sidereal day, solar day, and synodic month -- complicate the radial velocity periodogram yet allow us to break the degeneracy between alias and physical frequency that would exist for evenly-sampled data.

In the following section, we describe the origin and characteristics of aliases, supply the details of our approach for confirming that a particular frequency is not an alias, and clarify previous misconceptions about aliases. In the third section, we apply our approach to confirm periods for the planets GJ~876~d and HD~75898~b. We find that the orbital period for Gl~581~d and for the planets of HD~73526 cannot be definitively determined due to noise. We discover that the reported orbital period for HD~156668~b, 4.6455 days, is an alias of the true period,  1.2699 days. Finally, we analyze the 5-planet system 55 Cnc. We find that the period of 2.817 days reported in the literature for planet e \citep{2004M, 2008F} is actually a daily alias of its true period of 0.737 days. We conclude by summarizing the approach we have developed, considering the implications of a new period for 55~Cnc~e, and suggesting observational strategies for mitigating aliases.

\section{Method}
\label{sec:alias}

The existence of a planet orbiting a star is frequently inferred from a signature peak in the periodogram of radial velocity measurements of the star. However, the periodogram often contains alias frequencies, the result of discrete sampling times, that, at first glance, cannot be distinguished from the true periodicities. Many astronomers have struggled to determine which periodogram peaks are physical frequencies and which are aliases, often resorting to methods that are unnecessarily computationally intensive, not definitive, reflect a misunderstanding of aliases, or all of the above. In the first subsection, we will describe the origin of aliases for evenly and unevenly sampled data. In the second subsection, we will explain the cause of the daily aliases, prominent for many Doppler datasets. In the third section, we will present a field guide for identifying aliases. In the fourth section, we will describe the method we have developed. In the fifth subsection, we will discuss the effects of orbital eccentricity. In the sixth subsection, we will discuss common misconceptions about aliases that lead to misidentification. 

\subsection{The Origin of Aliases for Evenly and Unevenly Sampled Data}

Aliases are the result of discretely sampling a continuous signal. The resulting discretely-sampled signal is the product of the continuous signal and the sampling function, the latter being a ``Dirac comb'': a series of delta functions. The periodogram of the discretely-sampled signal is a convolution of periodogram of the continuous signal and the periodogram of the sampling function (the spectral window function). 
Consider first the simplified case of an infinite set of evenly-spaced data points, $g_1[n]$, the result of sampling a continuous sine wave $s_1(t)$ of frequency $f$ at sampling frequency $f_ s$. Here we follow \citet{1999M}:\\
\begin{eqnarray}
s_1(t) &=& \sin(2\pi f t), \nonumber \\
g_1[n] &=& s(n/f_s) = \sin(2\pi f n/f_s), \nonumber
\end{eqnarray} where $n$ is an integer. However, under this sampling, the signal is indistinguishable from the sine wave $s_2(t)$ of frequency $(f+m f_s)$:
\begin{eqnarray}
s_2(t) &=&  \sin(2\pi (f+m f_s) t), \nonumber \\
g_2[n] &=& \sin(2\pi(f+mf_s)n/f_s) =  \sin(2\pi f n/f_s), \nonumber 
\end{eqnarray} where $m$ is an integer. In the frequency domain, both $g_1$ and $g_2$ will have peaks not only at $f$, but also at $f+mf_s$.\\
Moreover, neither has a periodogram distinguishable from a sampled sinusoid of frequency $(-f+m f_s)$:
\begin{eqnarray}
g_3[n] &=& \sin(2\pi(-f+mf_s)n/f_s) = \sin(2\pi f n/f_s+\pi). \nonumber
\end{eqnarray}
That is to say, $g_1$ and $g_2$ will also have peaks at $-f+mf_s$, although the phase of those peaks will be advanced by $\onehalf$ cycle.  For evenly sampled data,  unless the only physically possible frequencies fall in a single Nyquist interval $f_s/2$, the frequency cannot be unambiguously determined.

\begin{figure}[htbp]
   \centering
   \includegraphics{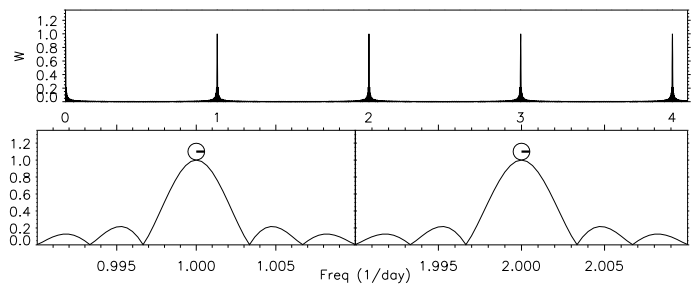}
   \caption{Spectral window of data evenly sampled in time, with a sampling frequency $f_s = 1$~day$^{-1}$, and 300 samples.
   \vspace{0.1 in}}
   \label{fig:wineven}
\end{figure}

Fig.~\ref{fig:wineven} shows the spectral window function of an evenly sampled time series of $f_s$ = 1 day$^{-1}$. Peaks in the spectral window function occur at $mf_s$, where $m$ is an integer.  The spectral window function is given by equation 8 in \citet{1987R}: 
\begin{equation}
W(\nu) = \frac{1}{N}\sum^N_{r=1} e^{-2 \pi i \nu t_r}, \label{eqn:wind}
\end{equation}
where $N$ is the number of data points, and $t_r$ is the time of the $r$th data point. It is evident that when $\nu = \pm m f_s$, $e^{\mp 2 \pi i m f_s t_r} = e^{\mp 2 \pi i m n} = 1$ and $W(\nu) = 1$. It's also evident from this equation that when $\nu=0$, $W(\nu) = 1$.  Note that $W(-\nu) = W^*(\nu)$.

\begin{figure*}[htbp]
   \centering
   \includegraphics{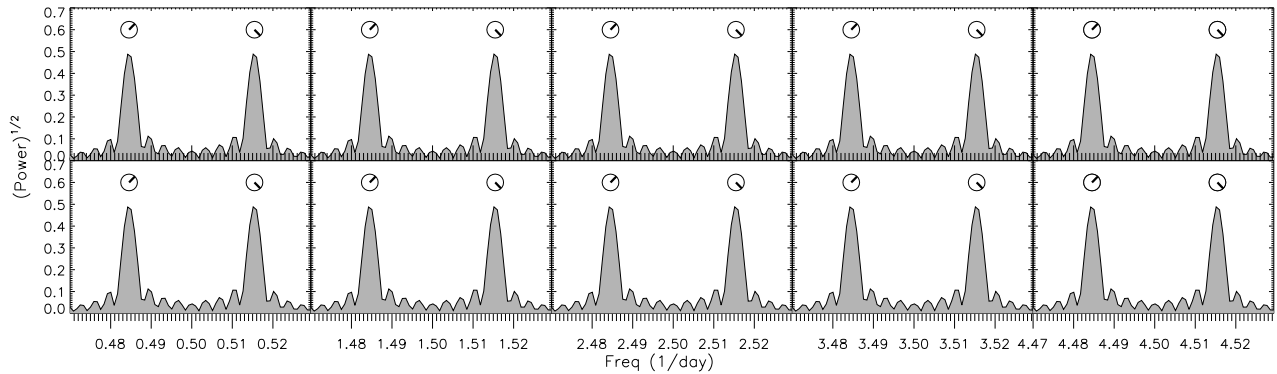}
   \caption{Periodogram of sinusoids sampled evenly in time.  Top:  Period 1.94 day. Bottom: Period 2.06 day. They are indistinguishable.
   \vspace{0.1 in}}
   \label{fig:speceven}
\end{figure*}

The top panel of Fig.~\ref{fig:speceven} shows the periodogram\footnote{For this and all other periodograms in this paper, at each frequency we (1) let the mean of the data float, and (2) weighted each data point with the inverse of the square of the reported error bar.  See \cite{1999C} and \cite{2009ZK}.} of a sinusoid of period 1.94 days sampled every 1 day for 300 days.  For a sinusoidal signal, the resulting periodogram is a convolution of the spectral window function $W(\nu)$ with the peak corresponding to period 1.94 days. The bottom panel shows the periodogram of a sinusoid of period 2.06 days, an alias of 1.94 days, with the same even sampling. The two periodograms are indistinguishable. The aliases of the 1.94 day period occur at $ f = 1/1.94 + m f_s$. For $f_s = 1$ and $m=-1$, the alias is $1/1.94 - 1 = 1/2.06$.

For a randomly selected frequency $\nu$ each $e^{-2 \pi i \nu t_r}$ will add incoherently. However, if there are gaps in the data of a certain frequency $\nu$, only certain phases occur and the complex exponentials will add in a partially coherent manner. The spectral window functions of stellar reflex motion measurements contain peaks at 1 sidereal year, 1 sidereal day, 1 solar day, and sometimes 1 synodic month. These periodicities are caused by observations being limited to only a particular portion of each of these periods. Observations are limited to a particular portion of the sidereal year and sidereal day because the star is only visible at night from the location of the telescope during particular parts of the sidereal year and day.  At some telescopes, spectroscopic observations of the stars are relegated to ``bright time," the portion of the synodic month when the moon is near full, because ``dark time" is reserved for observing faint objects.  In the next section, we will focus on the daily aliases due to both the solar day and the sidereal day.

Uneven sampling also dictates that the phase of $\exp(2 \pi i f_s t_r)$ will span a width. \citet{1999E} demonstrate that for unevenly sampled data, there is effectively no Nyquist frequency. Because gaps in the data and uneven spacing sample a non-zero width in phase, the height of peaks in the window function will never be exactly 1. For a noiseless data set, the physical frequency will almost always be a higher peak in the periodogram than any alias.  (The only exception is if positive and negative aliases add coherently.) For noisy data, the noise between two candidate peaks is correlated, but it may constructively interfere with the alias and destructively interfere with the true frequency, resulting in the alias peak being taller. Depending the phase of noise, it can also alter the phase of the true frequency and aliases through vector addition.

\subsection{Daily Aliases} \label{sec:daily}

For most Doppler datasets, the largest peaks in the window function --- corresponding to the largest aliases --- are those at $n$ day$^{-1}$, where $n$ is an integer.  We refer to these peaks as the daily aliases, as they result from the sampling an Earth-bound observer is able to do at nighttime from a single site.  

\begin{figure}
\includegraphics[scale=0.7]{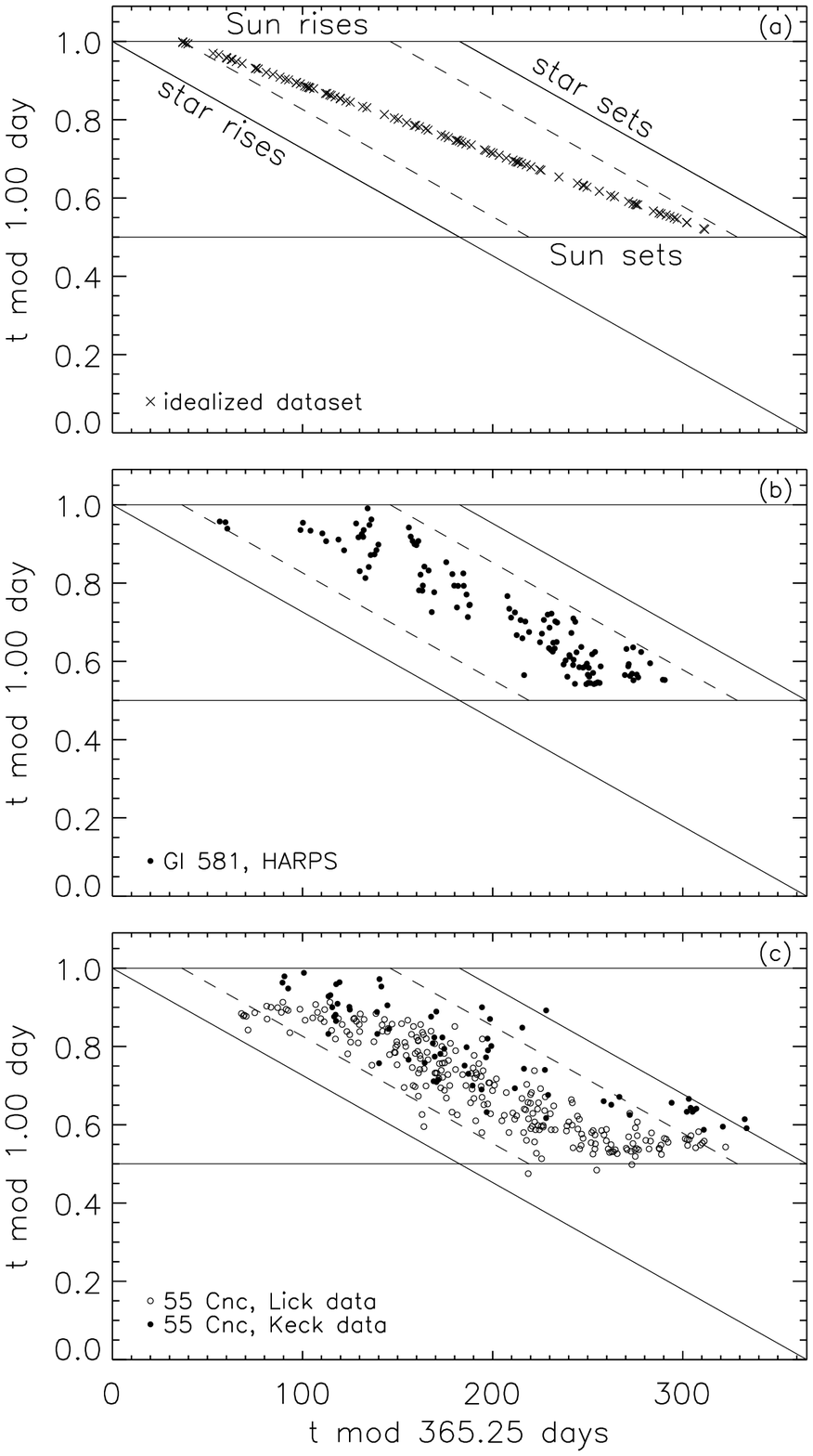}
\caption{Times of observation of an idealized dataset and two real datasets, folded to illustrate the origin of daily aliases.  The axes show, quantitatively, the time of the year and the time of the day.  The solid lines are labeled and correspond to the time each day that either the Sun rises or sets (at a constant time-of-day in this idealized example) or the star rises or sets (which varies according to the time of the year).  The dashed lines are when the star reaches 54 degrees from the zenith, within which a favorable observation can be made. The idealized dataset is described in the text.  The HARPS data for Gl~581 are from \cite{2009M}, and we took $t=JD-2,452,970.92$ for convenience.    The Keck and Lick data for 55 Cnc are from \cite{2008F}, and we took $t=JD-2,447,370.15$.   
\vspace{0.1 in}}
\label{fig:mw1}
\end{figure}

Let us construct an example dataset, to illustrate their origin.  Suppose the sampling is confined to when the Sun is down and the target star is up.  In particular, suppose the samples are taken nearly daily, midway between when the star rises and the Sun rises, or midway between when the Sun sets and the star sets, depending on the time of the year.  This sampling would lead to spacings between the solar day (24h 0m 0s) and the sidereal day (23h 56m 4s).  Therefore, in our example dataset, let us take datapoints spaced by 23h 57m 30s, although due to telescope scheduling and weather, only a fraction of the nights (randomly chosen) are actually observed.  Such a sequence is repeated in intervals of 365 days for 5 years, resulting in a total of 97 observation times.  In Fig.~\ref{fig:mw1} we illustrate this idealized dataset.  It is constructed to obey the boundaries set by the Sun and the star, which are also plotted.  The actual times from real datasets are compared, to show that this sampling, though idealized, reproduces the main daily and yearly structure of a real dataset.  

\begin{figure}[htbp]
   \centering
   \includegraphics{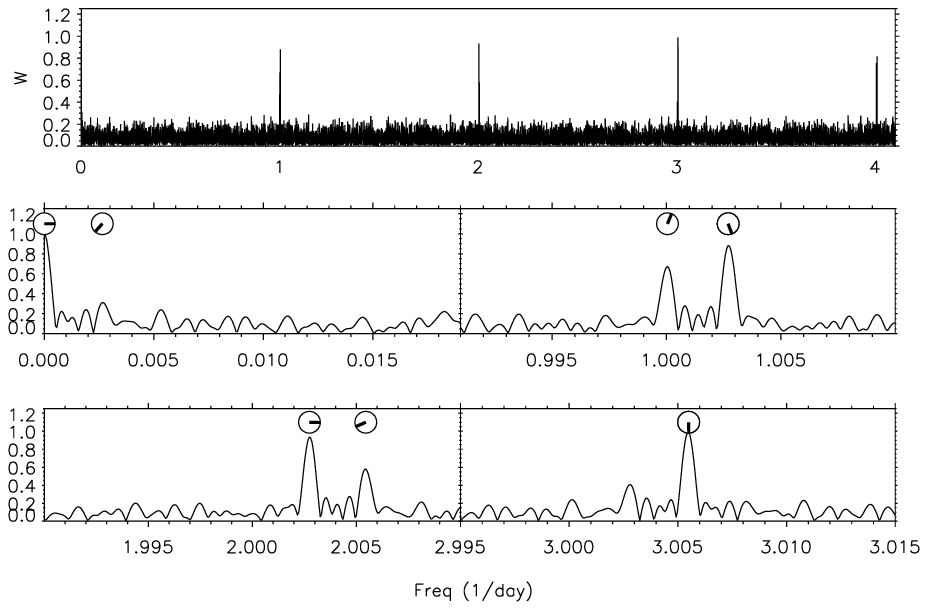}
   \caption{Spectral window function of data with gaps.  The sampling is from the ``idealized dataset'' of panel (a) in Fig.~\ref{fig:mw1}.}
   \label{fig:winfake}
\end{figure}

The window function for this idealized dataset is shown in Fig.~\ref{fig:winfake}.  There are peaks at frequencies of $n$~day$^{-1}$ + $m$~yr$^{-1}$.   In particular, there is a doublet at $\nu = 1.0000$~day$^{-1}$ and $\nu = 1.0027$~day$^{-1}$, with the latter peak being larger.

How does this structure arise?  We see from Fig.~\ref{fig:mw1} that for $\nu = 1.0000$~day$^{-1}$, the idealized observations only sample the second half of phase.  Therefore the window function as defined by equation~(\ref{eqn:wind}) will have contributions only from phases $\pi$ to $2 \pi$, so the complex exponential will add up coherently to a large peak.  This phase coherence explains  the daily aliases not just at $1$~day$^{-1}$, but everywhere a peak occurs.  For instance, consider at what times data are taken relative to the frequency of the sidereal day,  $\nu = 1.0027$~day$^{-1}$.  In Fig.~\ref{fig:mw1}, this frequency is related to the diagonal line labeled ``star rises.''  The idealized dataset consists only of observation times between $0.1$~days and $0.4$~days after the star rises (above that diagonal line).  Therefore the observations cover only $30\%$ of the phase of the sidereal sampling frequency, which again results in a large peak in the window function.  Here, even a smaller fraction of the total phase is covered, so the sampling results in even more coherent summation of complex exponentials, which is why the window function peak at $\nu = 1.0027$~day$^{-1}$ is larger than at $\nu = 1.0000$~day$^{-1}$ (Fig.~\ref{fig:winfake}).  Another way to see this is to note that the line formed by the idealized data in Fig.~\ref{fig:mw1}, panel a, has a slope more closely matching the sidereal day (the diagonal lines related to the star) than the solar day (the horizontal lines related to the Sun).  Finally, we note that no peaks in the window function appear between the solar and sidereal frequencies because folding the data at those frequencies samples phases throughout $0$ to $2\pi$.

Having understood the origin of the daily aliases in the window function, including doublets, we are prepared to recognize and correctly interpret such structure when it results in periodograms.

\begin{figure*}[htbp]
   \centering
   \includegraphics{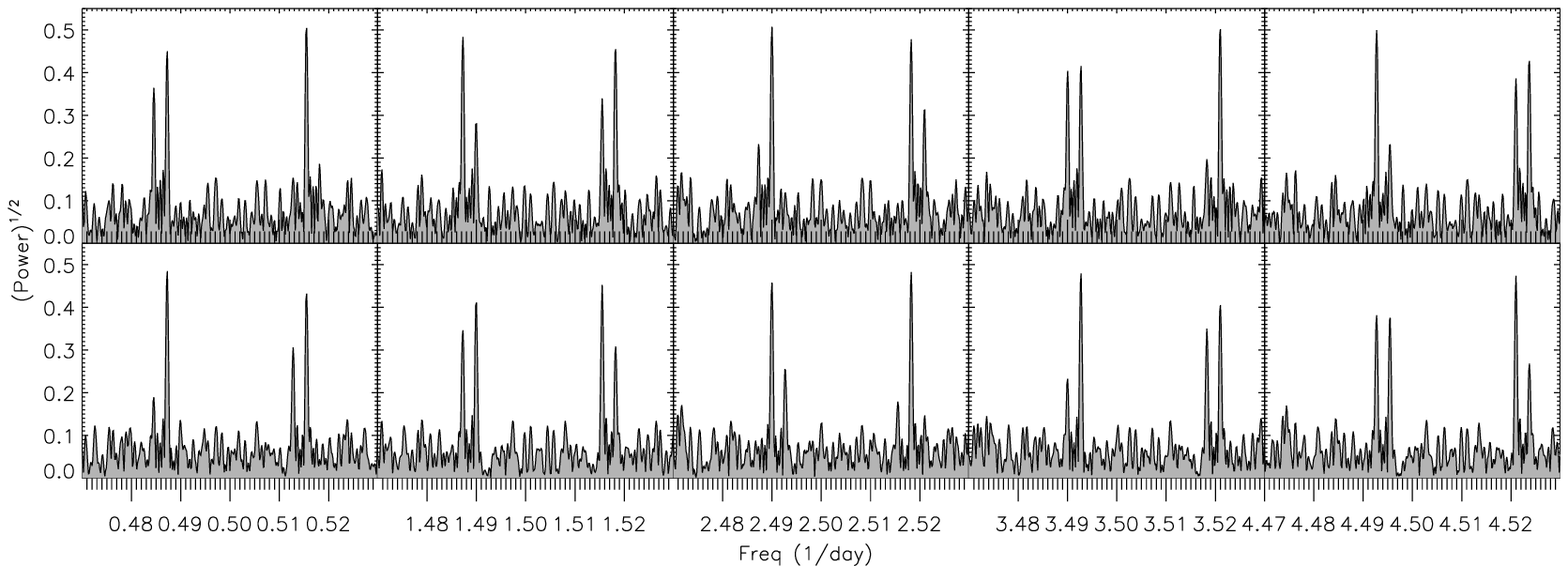}
   \caption{Top: Periodogram of sinusoid of period 1.94 day (frequency 0.515 day$^{-1}$) with the idealized time sampling from Fig.~\ref{fig:mw1}. Bottom: Period 2.06 day (frequency 0.485 day$^{-1}$). With this time sampling, the periods are distinguishable by the imprinting of the window function features from Fig.~\ref{fig:winfake} at $f \pm f_s$ where $f$ is the frequency of the sinusoid and $f_s$ of the window function feature.
   \vspace{0.1 in}}
   \label{fig:specfake}
\end{figure*}

To that end, we used this idealized dataset to sample a sinusoid of period 1.94 days or 2.06 days, and in Fig.~\ref{fig:specfake} show their periodograms. In this example, we have taken the two periods close to those which \cite{2005R} needed to decide between for GJ~876~d.  Here, then, we have identified a simple way to decide between them: the slightly taller peak is expected to be the true one (because there is no noise), and the alias will consist of a doublet with spacing $0.0027$~day$^{-1}$. We analyze the \citet{2005R} dataset in subsection \ref{subsec:GJ876}.

\subsection{A Field Guide to Aliases}

\begin{figure*}
\includegraphics{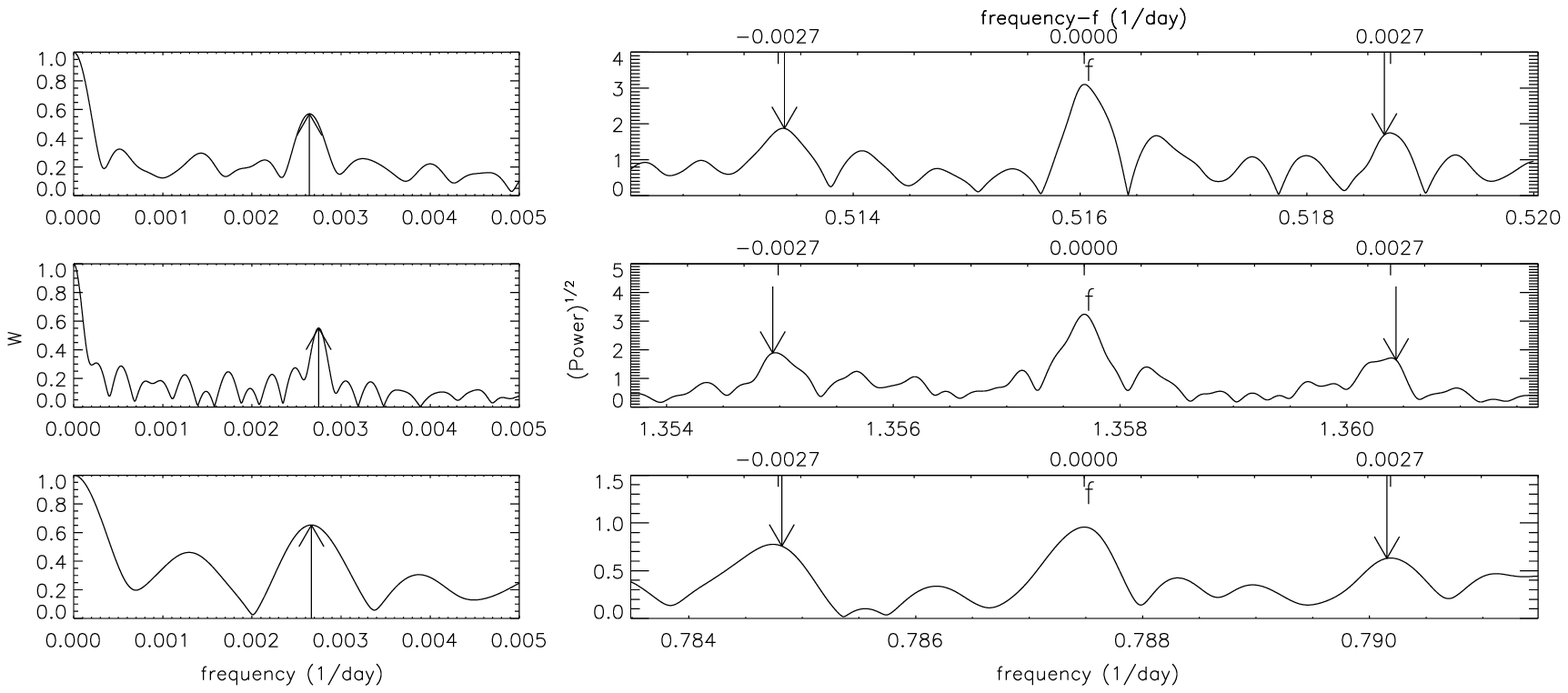}
\caption{Illustrative examples of yearly aliases taken from GJ~876 (top), 55 Cnc (middle), and HD~156668 (bottom). The window function is plotted on the left and the periodogram of the data near the candidate frequency on the right. The arrow in the left plots indicates the peak in the window function near 1/yr and the arrows in the right plots indicate the predicted location of the yearly aliases caused by this window function feature.
   \vspace{0.1 in}}
\label{fig:fgyr}
\end{figure*}

\begin{figure*}
\includegraphics{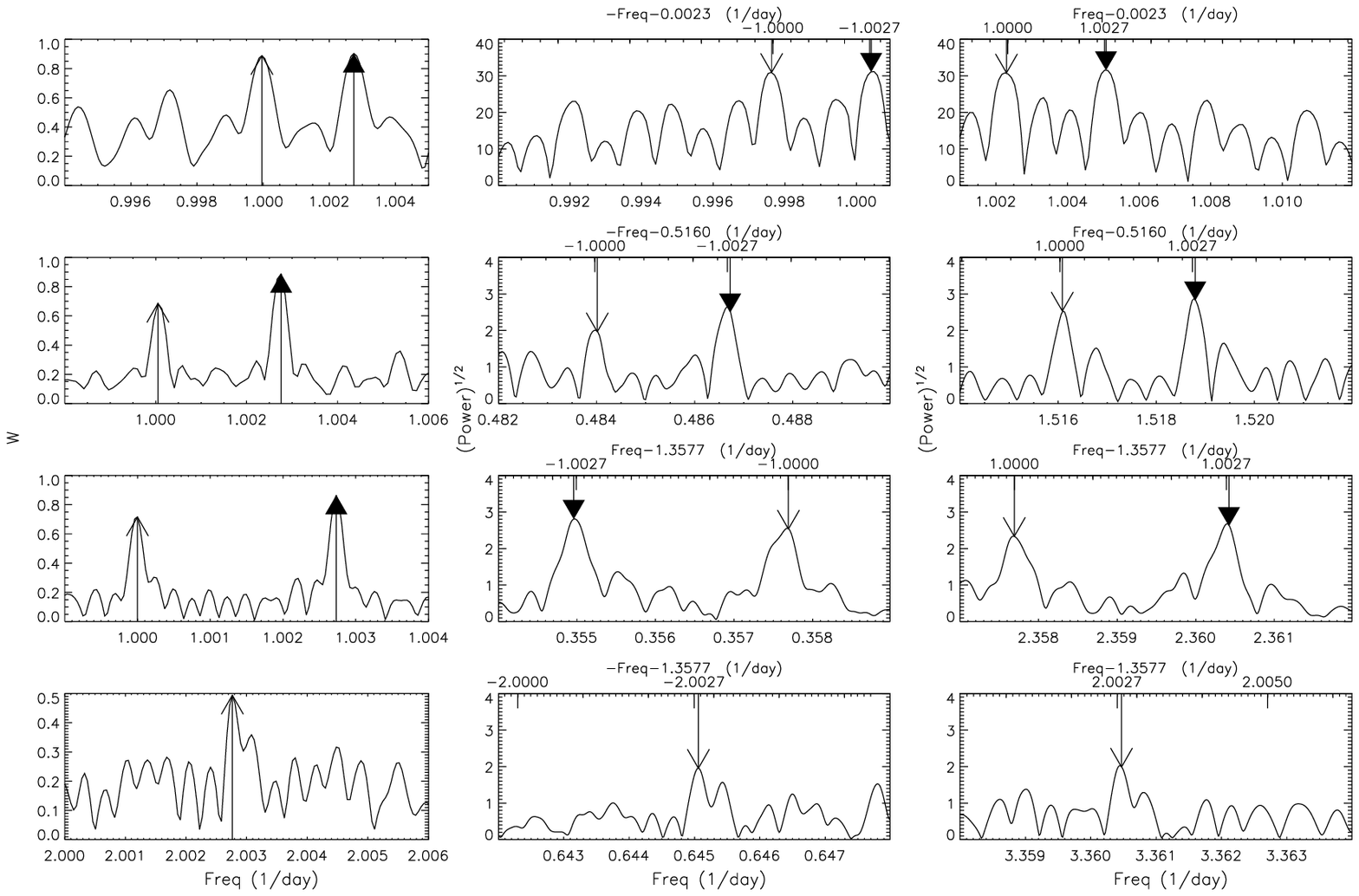}
\caption{Illustrative examples of yearly aliases taken from HD~75898 (top), GJ~876 (second row), 55 Cnc Fischer et al. data set (third row), and 55  Cnc combined data set (bottom row). The window function near a major feature is plotted in the left column and sections of the periodogram of the data in the middle and right columns. Arrows in the left column indicate the peaks in the window function near sidereal and solar days and the arrows in the middle and right plots indicate the predicted locations of the corresponding aliases. Note that each peak in the window function results in two features in the data periodogram.
   \vspace{0.1 in}}
\label{fig:fgday}
\end{figure*}

An alias is a convolution in frequency space of a physical frequency with the window function. Fig.~\ref{fig:fgyr} and Fig.~\ref{fig:fgday} display some examples of yearly and daily aliases respectively and the window function features that cause them. We have chosen especially clean examples; ambiguous cases will be addressed throughout the next section. 

\subsection{Details of Our Method}

We recommend the following treatment for a radial velocity dataset or residuals of an established fit (we will refer to both these categories as ``data") that appear to exhibit periodic variation. As we emphasized above, the phases of peaks are helpful for determining what is the true frequency. For example, consider a set of data with peaks in the spectral window function at 1 year (0.0027 day$^{-1}$), 1 solar day (1 day$^{-1}$), and 1 sidereal day (1.0027 day$^{-1}$). Consider a true frequency $f_1 > 1.0027$, which will have aliases at at $f_2 = f_1 - 1$ and $f_3 = f_1 -1.0027$. We may wonder if the peak at $f_2$ is the true frequency, with an alias at $f_s - 0.0027 = f_3$. However, because of the phases of the peaks in the window function, the phase of the peak $f_3$ is different than the phase we would expect if it were an alias of $f_2$. 
Because the phase of a peak can be key in determining the true frequency, we strongly recommend plotting the phase of selected peaks. We use a symbol we call a ``dial" (e.g., Fig.~\ref{fig:gj876}) where the phase angle is the counterclockwise angular position from the x-axis. The phase angle is $\tan^{-1}($Imaginary$(W(\nu))/$Real$(W(\nu)))$ for the window function peaks and likewise $\tan^{-1}(C(f)/B(f))$ for the periodogram peaks, where $B$ and $C$ are the real and imaginary coefficients of the periodogram for frequency $f$.

Our method is composed of the following steps:

\begin{enumerate}

\item Plot the spectral window function (eq.~\ref{eqn:wind}), attaching dials to any large peaks. Peaks will most likely occur at or near f=0, 1/yr, 1/(solar day), 1/(1 sidereal day), and, if the observations were taken during a particular part of the lunar cycle, f = 1/month, 1/month $\pm$ 1/yr. Spectral window functions of artificial data sets are plotted in Fig.~\ref{fig:wineven} and \ref{fig:winfake} and real data sets in Fig.~\ref{fig:wingj876}, \ref{fig:winhd75898}, \ref{fig:hd73526}, \ref{fig:wingl581}, \ref{fig:winhd156668}, \ref{fig:win55cnchet}, \ref{fig:win55cnccomb} and \ref{fig:win55cncfischer}.

\item Plot the periodogram.

\item Consider first the possibility that the largest peak is the true frequency; measure its frequency, phase, and amplitude. Attach dials to peaks we would expect are aliases, according to the peaks in the window function. If the peak in the radial velocity periodogram occurs at $f$ and peaks in the window function occur at $f_s$, we expect aliases at $f \pm f_s$. (If $f_s > f$, we will still see a peak at $f-f_s$. Flipping it across 0 frequency gives the phase the opposite sign: a complex conjugation.) Generate a sinusoid with the same frequency, phase, and amplitude as the peak in radial velocity periodogram and plot its periodogram, attaching dials in the same location. Compare the amplitude and phases of peaks. Are the major aliases for $f$ present in the data with the predicted phase and amplitude? 

\item Now consider that the largest alias(es) of what we considered the true frequency might actually be the true frequency. Repeat step 3.

\item If the periodogram of the data is well-matched by the periodogram of one and only one candidate sinusoid, then the true frequency has been determined. As \citet{1976L} said, ``If there is a satisfactory match between an observed spectrum and a noise-free spectrum of period $P$, then $P$ is the true period." However, if several candidate sinusoids match peaks equally well or poorly, then the data are not sufficient to distinguish the true period. 

\end{enumerate}

\subsection{Treating the Orbital Eccentricity}
Many extra-solar planets have elliptical orbits. The signal of the eccentricity is contained in harmonics of the orbital frequency; the first harmonic has an amplitude $eK$, where $e$ is the eccentricity and $K$ is the amplitude of the sine wave at the planet's orbital frequency \citep{2010A}. Thus for moderate eccentricities, the same analysis can be applied to the first harmonic of the orbital frequency. Except in rare unfortunate cases (such as HD~73526, treated below in section \ref{subsec:HD73526}), the period and its aliases will be well-separated in frequency space from the eccentricity harmonic and its aliases. In section \ref{subsec:HD75898}, we distinguish a peak in the periodogram of $HD~75898~b$ as an alias that the \citet{2007R} proposed could be an alias, eccentricity harmonic, or additional planet. 

For certain datasets, orbital eccentricity may help distinguish between a true orbital period and an alias. Consider a planet with moderate eccentricity $e$ whose host star is observed with near evenly-spaced sampling as $f_s$. Even if the noise is low relative to $K$, it may be difficult to distinguish between the true orbital frequency $f$ and an alias $f + f_s$. However, since the planet's orbit is eccentric, we will also observe a peak of amplitude $eK$ at $2f$ but no such peak at $2(f+f_s)$.

In summary, orbital eccentricity contributes to the periodogram in a well-defined way and, except in rare unfortunate cases that can be easily identified, will not confuse the distinction between the true orbital period and an alias.

\subsection{Common Misconceptions}

Many problems with aliases are the result of unwarranted assumptions. We describe some common misconceptions about aliases and how they cause confusion.

\begin{enumerate}

\item Assuming that the largest peak in the periodogram is the physical frequency. In fact, noise may add coherently to an alias or incoherently to the physical frequency, causing the alias to appear larger. This is what happened for Gl~581 \citep{2007U,2009M}.  In multi-planet systems, aliases from several planets could add to make the highest peak a spurious signal \citep{1995F}.

\item Assuming that the frequency that yields the best Keplerian or Newtonian planet fit is the true frequency. As we saw for Gl~581~d, this is not always the case, due to noise.

\item Assuming that aliases occur at frequencies only \emph{occur} near peaks in the spectral window function. We have seen authors plot the spectral window function below the periodogram of the data and assume that if a frequency in the data periodogram is not near a peak in the spectral window function that it is not an alias. In fact, aliases occur at $|f \pm f_s |$, where $f_s$ is a feature in the spectral window function. Depending on the relative values of $f$ and $f_s$, the alias might be anywhere in the periodogram. However, periodograms will contain peaks at the sampling frequencies if there are systemics linked with the observing pattern or if the peaks are aliases of a very low frequency signal. We emphasize the difference between these two types of signals: the former is spurious and and the latter has an extra-solar origin but wrong frequency. We also emphasize the importance of employing the spectral window function to identify all major aliases, not just aliases or other spurious frequencies that occur at the sampling frequencies.

\item Assuming that any frequency above 1 is an alias. As we mentioned above, there is effectively no Nyquist frequency for unevenly sampled data. Many authors cut off their periodograms at 1 day$^{-1}$, potentially missing out on or misinterpreting planets with orbital periods less than a day. We know such planets exist because they have been detected by transits. Moreover, because long period planets will have aliases near 1 day, a planet with orbital periods near 1 day is vulnerable to being discarded as an alias \citep{2007K}.

\item Assuming that aliases are so pernicious that one can never identify the correct period and should thus just pick the most sensible period. In fact, our method allows one to determine either a correct period or that noise prevents the identification of the correct period. In the latter case, further observations should allow for a definitive determination in the future. It is unwise to judge a priori which period is the most ``sensible" period; as mentioned above, planets have been found with periods less than a day.

\item Assuming that if an alias frequency is used in a Keplerian or Newtonian planet fit, a peak corresponding to the true frequency will appear in residuals. This would only happen if the peak at the alias frequency is much smaller than the peak at the true frequency, relative to the noise.

\item Assuming that if a frequency is an alias, it will appear in a periodogram of the data scrambled. Aliases are not caused solely by the spacing observations; they are convolution of the spectral window function with the periodogram of the data. Scrambling the data removes the true frequency and thus also removes the alias.

\item Assuming that if you ``fold" (i.e., phase) the data with a candidate period, a coherent pattern will emerge only if the candidate period is the physical period. In fact, a large alias, by its very definition, will also produce a coherent pattern.

\end{enumerate}

Another method we have seen applied to distinguish between two frequencies, one of which is an alias, is to generate thousands of mock data sets for each frequency by combining a sinusoid with simulated noise and then determine how often the alias is mistaken for the true frequency.  This method indeed reveals the probability that the period is falsely determined, but a proper understanding of the window function leads to a less computationally intensive method, which we have advocated.

We reemphasize the peaks in the spectral window functions combined with the true frequencies are what cause aliases. Even if a peak in the periodogram is linked to another peak by close to an integer frequency, if that integer frequency is not a peak in the spectral window function, then the peaks are not aliases of one another and might represent two distinct planets. Rather than simply noting the possibility that an integer frequency might link the peaks, the window function reveals it quantitatively.

\section{Application to Extrasolar Planetary Systems}

In the following section, we investigate instances of aliases and ambiguous periods in the literature.

\subsection{GJ~876~d}
\label{subsec:GJ876}
In this section, we apply the approach described above to planetary system GJ~876. Extensive radial velocity observations spanning almost eight years have revealed three planets orbiting this M-star. A Jupiter-mass planet $b$ was discovered in 1998 \citep{1998M}, and an interior Jupiter-mass planet $c$ in a 2:1 resonance with $b$ was discovered three years later \citep{2001M}. After several years of continued observations, \citet{2005R} discovered an additional 7.5 earth mass planet $d$ with an orbital period of 1.94 days. This discovery was independently confirmed by \citet{2010C} with new HARPS data. The periodogram of the residuals to the nominal two-planet, $i = 90^\circ$, coplanar fit exhibits strong power at frequency 0.52 day$^{-1}$ but also at $f = 0.49$ day$^{-1}$ ($P = 2.05$ day) and $f =1.52$ day$^{-1}$ ($P = 0.66$ day) (Fig.~\ref{fig:gj876}, top panel). \citet{2005R} performed a series of tests and argued based on the results that the peak at 2.05 days is an alias of the true period at 1.94 days. Our method is able to definitively confirm the results of their tests, that the physical period is indeed 1.94 days.

\begin{figure}[htbp]
   \centering
   \includegraphics{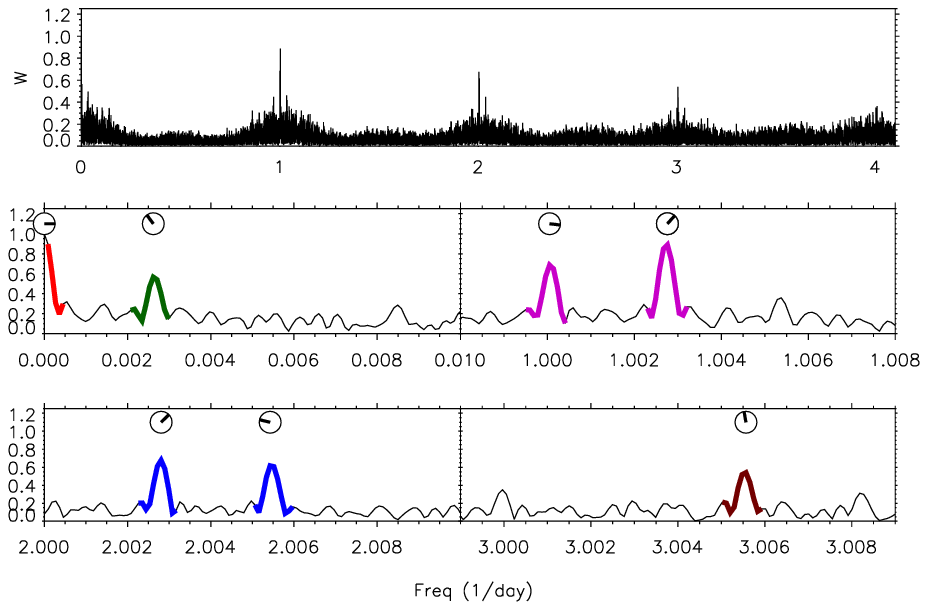}
   \caption{Spectral window function of RV Measurements of GJ~876 \citep{2005R}. Major features of the spectral window function are colored: red (at 0 day$^{-1}$), green (yearly feature), fuschia (daily features), blue (two day$^{-1}$), and brown (three day$^{-1}$). The corresponding aliases these features cause for several candidate frequencies are indicated by these colors in Fig.~\ref{fig:gj876}.
   \vspace{0.1 in}}
   \label{fig:wingj876}
\end{figure}

\begin{figure*}[htbp]
   \centering
   \includegraphics{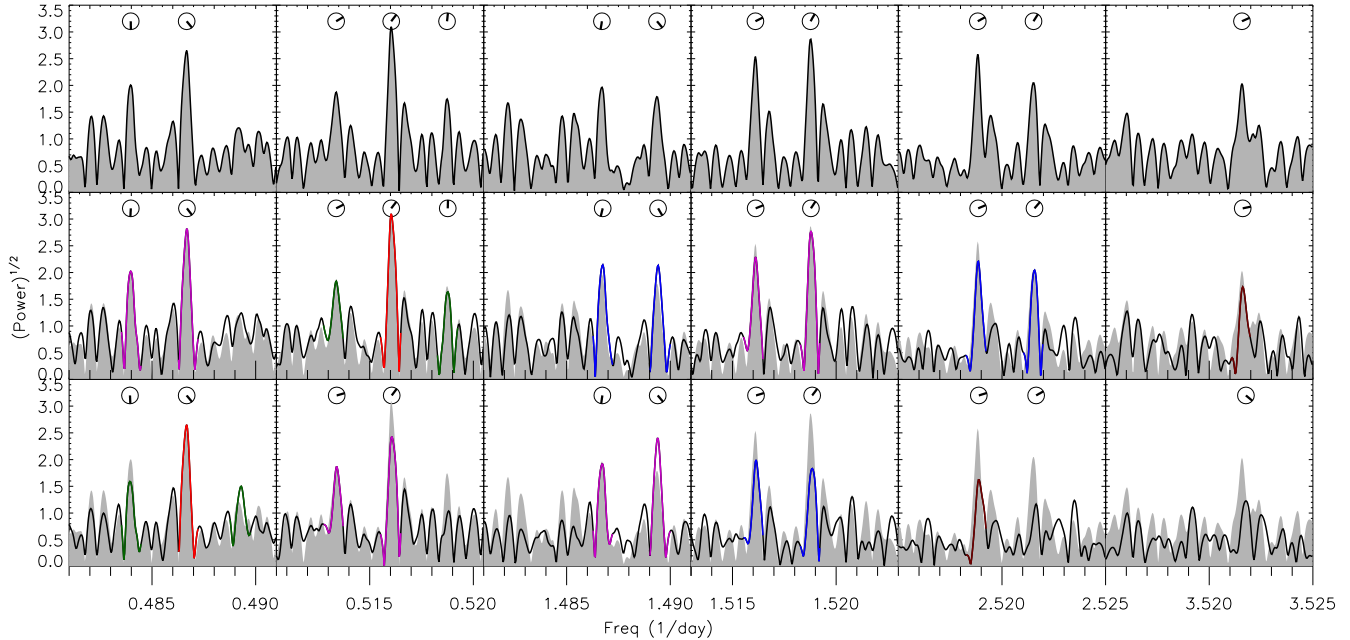}
   \caption{Periodograms of GJ~876.  The top row is the periodogram of the data. The second and third rows show the periodograms of
sinusoids sampled at the times of the real data sets as solid lines; they also repeat the periodogram of the data as a gray background, for comparison.  Dials above the peaks show the phase at each peak. Colors correspond to the feature in the window function that creates the particular alias (see Fig.~\ref{fig:wingj876}), with red being the candidate frequency, the green sidebands yearly aliases, and the fuschia, blue, and brown peaks daily, two day$^{-1}$, and three day$^{-1}$ aliases respectively. The second row is the periodogram of an injected sinusoid of period 1.94 days (frequency 0.516 day$^{-1}$).  The third row is the periodogram of an injected sinusoid of period 2.05 days (frequency 0.487 day$^{-1}$).  The sinusoid of period 1.94 days matches the heights and phases of the peaks much better, both for the yearly aliases on either side of the main peak in column 2 and the daily aliases in the other columns. The two candidate frequencies have different types of aliases at different locations, allowing us to break the degeneracy.
   \vspace{0.1 in}}
   \label{fig:gj876}
\end{figure*}

The spectral window function and the periodogram of GJ~876 (actually, of the residuals from a dynamical fit to planets b and c) are shown in Fig.~\ref{fig:wingj876} and \ref{fig:gj876}, respectively.  Major peaks in the window function occur at 1 sidereal year, 1 sidereal day, and 1 solar day.  The very same features are seen in the example periodogram described in section \ref{sec:daily}.  The main peak is tallest \footnote{We point this out for identification purposes but in a given data set, because of noise, the true frequency will not necessarily be taller than the alias.}.  The alias has a doublet structure.  

\clearpage
Compared to our example idealized data set of Fig.~\ref{fig:mw1}, \ref{fig:winfake}, and \ref{fig:specfake}, the yearly aliases are more pronounced in the data, because the observing season is shorter than in our idealized dataset.  This causes peaks on either side of the true peak (spaced by 1~yr$^{-1}$ = 0.0027 days$^{-1}$) which are symmetric in height.  Thus we confirm the selection of $P=1.94$~days as the correct period of GJ~876~d \citep{2005R}, and thus we demonstrate that a signal beyond the traditional Nyquist frequency can be robustly detected with unevenly sampled data.

\subsection{HD~75898~b}
\label{subsec:HD75898}

\begin{figure}[htbp]
   \centering
   \includegraphics[scale=0.9]{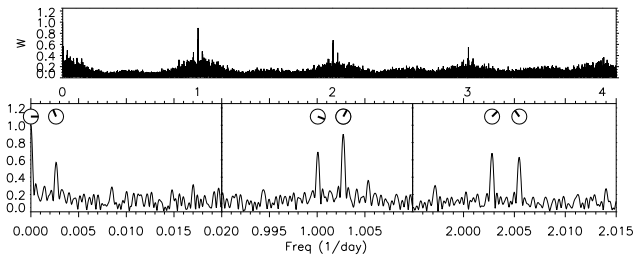}
   \caption{Spectral window function of RV Measurements of HD~75898. These features, convolved with a planet's orbital frequency, cause the aliases evident in the periodogram in Fig. \ref{fig:hd75898}.
   \vspace{0.1 in}}
   \label{fig:winhd75898}
\end{figure}

\begin{figure}[htbp]
   \centering
   \includegraphics[scale=0.9]{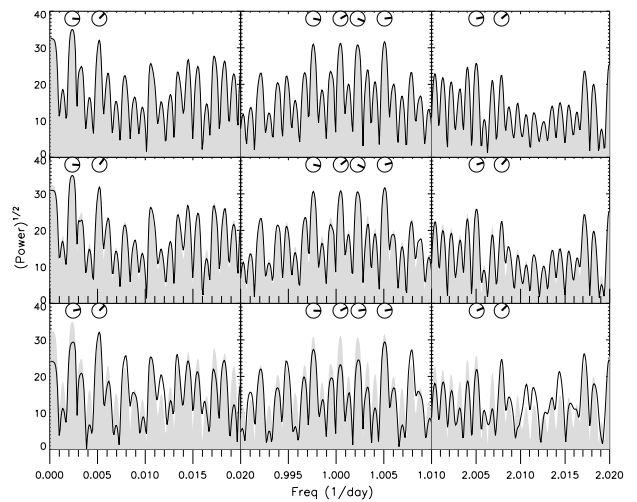}
   \caption{Periodograms of HD~75898. Dials above the peaks denote their phase.  Row 1 shows the data.  The other rows show sinusoids sampled at the times of the real data sets (solid line and dial), as well as the data again for reference (in gray).  In Row 2 the solid line shows, for these time samplings, the periodogram of a sinusoid of frequency 0.00236 day$^{-1}$.  For Row 3, it is for 0.00519 day$^{-1}$. We confirm that the peak at 0.00519 day$^{-1}$ is an alias, not a second planet or eccentricity harmonic. In Rows 2 and 3, each peak results from the convolution of the sinusoidal frequency with the features in the spectral window function in Fig. \ref{fig:winhd75898}.
   \vspace{0.1 in}}
   \label{fig:hd75898}
\end{figure}

\citet{2007R} discovered a Jupiter-mass planet orbiting HD~75898~b. They noticed two peaks in the periodogram, a large one near 400 days and a smaller one near 200 days. They presented three possibilities for the peak near 200 days: an alias of the 400 day period, an eccentricity harmonic (which we would indeed expect to appear near $P/2 = 200$ days), or a second planet. Applying our method, we confirm that the true period is 400 days, not 200 days; and the peak at 200 days is indeed an alias, not an eccentricity harmonic or second planet. The spectral window function is plotted in Fig.~\ref{fig:winhd75898}; the peak that occurs at 1 yr$^{-1}$ is the cause of the 200 day alias. In Fig.~\ref{fig:hd75898}, the periodogram shows that a 400 day period (row 2) produces exactly the aliases we expect, including the alias at 200 days. Although an eccentricity harmonic would fall at the same place as this alias, for this system we can rule out a significant eccentricity harmonic because the peak has the exact phase and amplitude that result from it being an alias of the 400 day planet; any significant eccentricity harmonic would change the phase and/or amplitude of this peak. These plots also confirm that the true period is 400 days, not 200 days (row 3).

\subsection{HD~73526}
\label{subsec:HD73526}

\begin{figure}[htbp]
   \centering
   \includegraphics[scale=0.9]{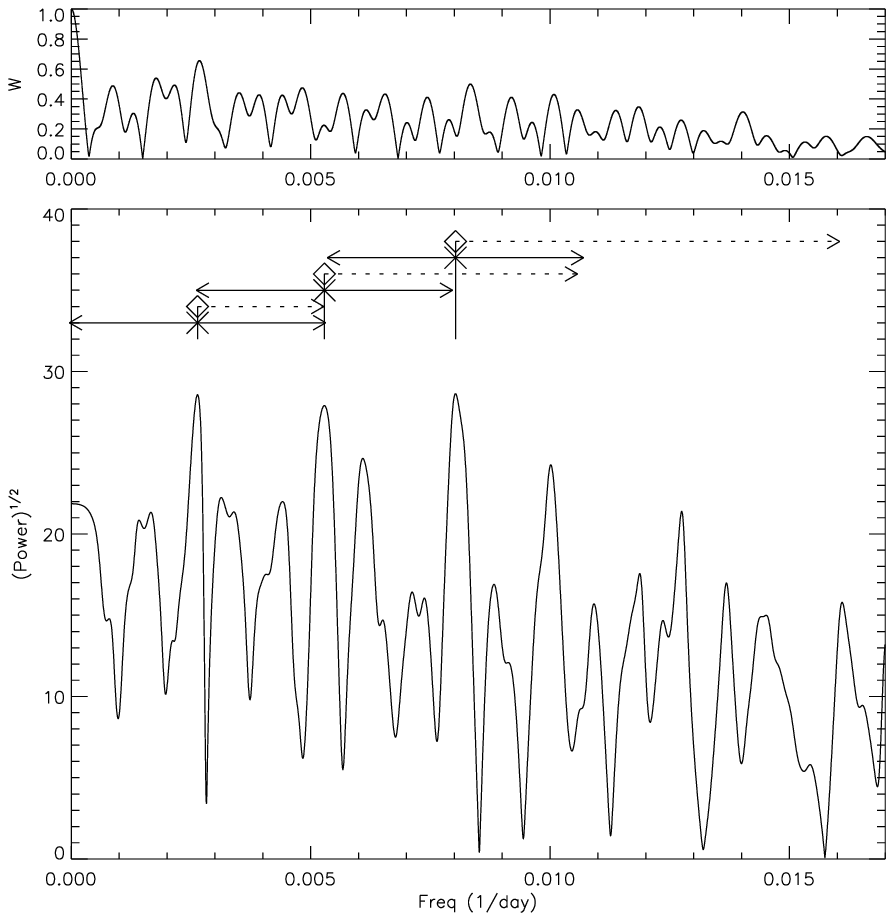}
   \caption{Top Panel: Spectral window function of RV measurements of HD~73526. Bottom Panel: Periodogram of RV measurements of HD~73526. The solid arrows indicate the locations of a peak's yearly aliases and the dashed line the location of the eccentricity harmonic.  \vspace{0.1 in}}
   \label{fig:hd73526}
\end{figure}

\citet{2003T} reported a planet orbiting the G-type star HD~73526 with orbital period 190.5 days. A later Bayesian analysis by \citet{2005G} revealed three possible periods for the planet: 190.4 days and (its yearly aliases) 127.88 days and 376.2 days. \citet{2005G} concluded that the periods 127.88 days and 376.2 days were more probable. After follow-up observations, \citet{2006T} reported the system actually contained two planets, with orbital periods 187.5 and 376.9 days, locked in a 2:1 resonance. The Keplerian fit using these two periods is an excellent match to the data, with $(\chi_\nu^2)^{1/2}=1.09$, but the dynamical fit for the system is substantially worse, with $(\chi_\nu^2)^{1/2}=1.57$. This implies that, though these periodicities may be strongly present in the system, the physical model of two planets orbiting with this period may need modification. Further complicating the interpretation of the system's periodicities is the degeneracy between the outer planet's eccentricity and the inner planet's mass -- or even its very existence \citep{2010A}. The window function for this system and a periodogram is plotted in Fig.~\ref{fig:hd73526}. The Keplerian fit has eccentricities of 0.4 for both planets, essentially tuning the phase of the power at 187.5 days (the first eccentricity harmonic of 376.9 days) to account for both a possible planet there and aliasing from 376.9 days; and introducing power at 93.8 days (the first eccentricity harmonic of 187.5 days and also a yearly alias of 127 days). However the eccentricities for the dynamical fit \citep{2006T} are substantially lower, implying that high eccentricities would cause dynamical interactions inconsistent with the data. It is possible that the periods 127.88 days and 376.2 days are incorrect but that by introducing a large eccentricity harmonic, the combination of orbital periods, eccentricity harmonics, and aliases match the periodicities of the data, which may be the result of different physical orbital frequencies. This system is complicated because of the degeneracy in frequency between resonant planets, eccentricity, and aliases. We recommend further observations and modeling of this system to confirm the orbital periods.  

\subsection{Gl~581~d}

\begin{figure}[htbp]
   \centering
   \includegraphics[scale=0.9]{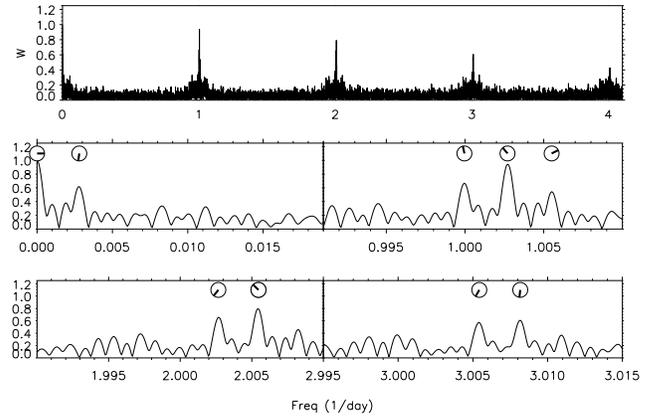}
   \caption{Spectral window function of Gl~581. These features, convolved with a planet's orbital frequency, cause the aliases evident in the periodogram in Fig. \ref{fig:gl581}.
   \vspace{0.1 in}}
  \label{fig:wingl581}
\end{figure}

\begin{figure*}[htbp]
   \centering
   \includegraphics{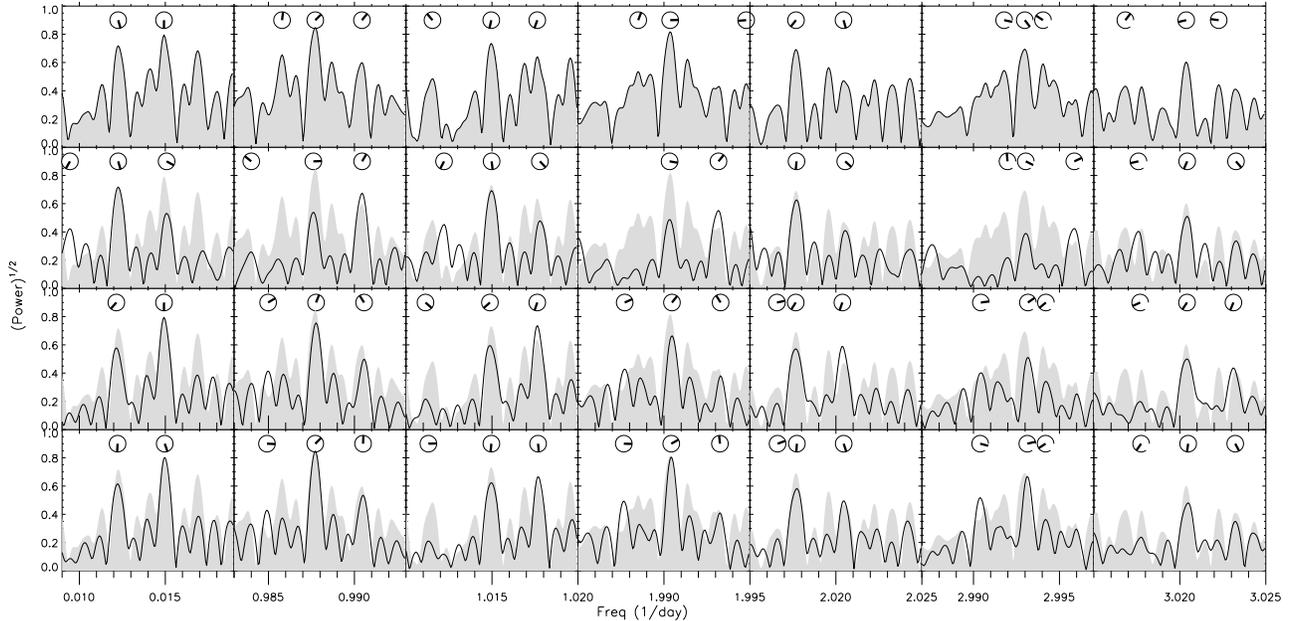}
   \caption{Periodograms of Gl~581 for planet d (planets b and c subtracted have been removed from the data set and planet e has been ignored; we obtain consistent results if we also remove planet e).  Dials above the peaks denote their phase.  Row 1 shows the data.  The other rows show sinusoids sampled at the times of the real data sets (solid line and dial), as well as the data again for reference (in gray). In Row 2 the solid line shows, for these time samplings, the periodogram of a sinusoid of frequency 0.0122 day$^{-1}$.  For Row 3, it is for 0.0150 day$^{-1}$. For Row 4, 0.9877 day$^{-1}$.  In Rows 2-4, each peak results from the convolution of the sinusoidal frequency with the features in the spectral window function in Fig. \ref{fig:wingl581}. Note that the phases and amplitudes of 0.0122 day$^{-1}$ and  0.0150 day$^{-1}$ are not consistent with the aliases we would expect. The period remains ambiguous, but we favor 0.9877 day$^{-1}$ based on this data set.  \vspace{0.1 in}}
   \label{fig:gl581}
\end{figure*}

HARPS measurements have revealed four planets orbiting the M dwarf Gl~581: a $\sim 2 M_\oplus$ planet e \citep{2009M}, Neptune-mass planet b \citep{2005B}, and super-Earth planets c and d \citep{2007U}. Planet d was originally reported to have a period of 83 days, beyond the cold edge of the habitable zone. After further observations, the HARPS team announced that the true period of planet d is 67 days, placing it within the habitable zone, and that the original 83 day period was a one year alias of the true 67 day period. In Fig.~\ref{fig:wingl581}, we plot the spectral window function of  \citet{2009M}'s new data set. Prominent peaks are evident at 1 year, 1 sidereal day, and 1 solar day.  A periodogram of the data, with planets b and c subtracted (subtracting planet e made no significant difference) and sinusoids of several candidate frequencies are plotted in Fig.~\ref{fig:gl581}. In the original data set, the highest peak in the periodogram was at 0.0122 day$^{-1}$ (corresponding to a period of 83 days). In the new data set, the highest peak is at 0.9877 day$^{-1}$. The second highest peak is at 0.0150 day$^{-1}$ (67 days), the period reported by \citet{2009M}. The 0.0122 day$^{-1}$ peak and 0.0150 day$^{-1}$ are linked by a feature in the window function at 1 sidereal year. Yet neither produces an alias that corresponds to the other frequency with a phase and amplitude that match the data (first column of rows 2 and 3). The highest peak, 0.9877 day$^{-1}$, is linked to the peaks at 0.0122 day$^{-1}$ and 0.0150 days$^{-1}$ by the window function peaks at 1 solar day and 1 sidereal day respectively; it better matches the phase and amplitude at these frequencies (row 4, column 1). This dataset has sampling which is too regular (Fig.~\ref{fig:mw1}b), which resulted in pernicious daily aliases. However, there are discrepancies between the phase and amplitude of the aliases predicted by all three candidate frequencies. For example, at 1.99 days (column 4), the larger alias predicted for  0.9877 day$^{-1}$ (linked by the large 1 sidereal day alias) is consistent in amplitude with the data while the other frequencies (linked by the smaller window function feature at 2 days) predict aliases that are too small; however, the phase for the 0.9877 day$^{-1}$ alias is a bit off. Although none of the frequencies is fully consistent, we slightly prefer 0.9877 day$^{-1}$, followed by 0.0150 day$^{-1}$ and 0.0122 day$^{-1}$. However, using the previous data set from \citet{2007U}, we favor (in order): 0.0122 day$^{-1}$, 0.0150 day$^{-1}$, and 0.9877 day$^{-1}$. We also fit a four-planet Keplerian model to both datasets. In the \citet{2007U} dataset, a frequency of 0.0122 day$^{-1}$ for planet d gave the best fit, while in the \citet{2009M} dataset, a period of 0.9877 day$^{-1}$ gave the best fit. However, a model with orbital frequency 0.0122 day$^{-1}$ where $e_d$ is allowed to float gives a significantly better fit than one with orbital frequency 0.9877 day$^{-1}$ where $e_d$ is fixed at zero (which would likely be attained by tidal dissipation). Because the period of planet $d$ remains ambiguous, we recommend that future observations take place with the star at a greater air mass -- instead of only when the star is crossing the meridian -- in order to reduce the amplitude of the aliases and allow us to definitively distinguish between these three candidate periods.

\subsection{HD~156668~b}
\begin{figure}[htbp]
   \centering
   \includegraphics{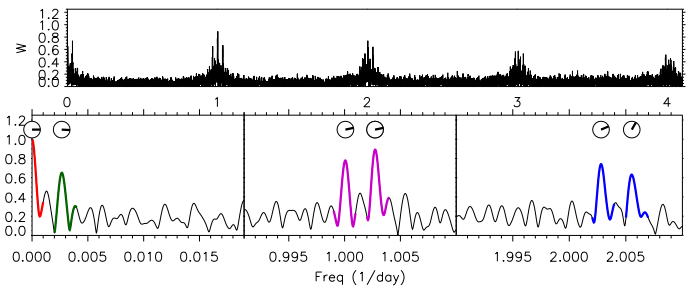}
   \caption{Spectral window function of RV measurements of HD~156668. Major features of the spectral window function are colored: red (at 0 day$^{-1}$), green (yearly feature), fuschia (daily features), and blue (two day$^{-1}$). The corresponding aliases these features cause for several candidate frequencies are indicated by these colors in \ref{fig:hd156668}.
   \vspace{0.1 in}}
   \label{fig:winhd156668}
\end{figure}

\begin{figure}[htbp]
   \centering
   \includegraphics{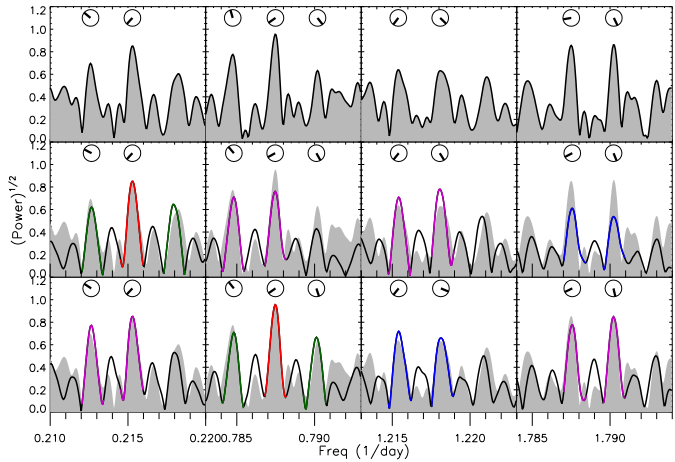}
   \caption{Periodograms of HD~156668. Row 1 shows the data. The other rows show sinusoids sampled at the times of the real data sets (solid line and dial), as well as the data again for reference (in gray). Colors correspond to the feature in the window function that creates the particular alias (see Fig.~\ref{fig:winhd156668}),  with red being the candidate frequency, the green sidebands yearly aliases, and the fuschia and blue peaks daily and two day$^{-1}$ aliases respectively. In Row 2 the solid line shows, for these time samplings, the periodogram of a sinusoid of frequency 0.215 day$^{-1}$.  For Row 3, it is for 0.787 day$^{-1}$, our favored value. The two candidate frequencies have different types of aliases at different locations, allowing us to break the degeneracy.
   \vspace{0.1 in}}
   \label{fig:hd156668}
\end{figure}

\citet{2010H} reported a 4 M$_\Earth$ planet orbiting HD~156668~b with period 4.6455 days (a frequency of 0.2153 day$^{-1}$). However, they considered that the correct period might be 1.2699 days (a frequency of 0.7875 day$^{-1}$), and our analysis confirms that as the correct period, as follows. The window function for this system is plotted in Fig.~\ref{fig:winhd156668} and periodograms of the data and sinusoids at two candidate frequencies in Fig.~\ref{fig:hd156668}. Note that large peaks in the window function occur at 1 sidereal and 1 synodic day while smaller peaks occur near 2 days (Fig.~\ref{fig:winhd156668}). For a true frequency of  0.2153 day$^{-1}$ (second row), we would expect two pairs of large peaks due to sidereal and solar aliases (second row, second and third column) and a smaller pair of peaks for the $\sim$2  day$^{-1}$ aliases (second row, fourth column). On the other hand, for a true frequency of 0.7875 day$^{-1}$ (third row), we would expect two pairs of large peaks due to sidereal and solar aliases (third row, first and fourth column) and a smaller pair of peaks for the $\sim$ 2 day$^{-1}$ aliases (third row, third column). The phase and amplitude of these aliases predicted for 0.7875 day$^{-1}$ (row 3) are thus more consistent with the data (row 1). Therefore we conclude that the planet's true period is 1.2699 days and that the peak at period 4.6455 days identified by \citet{2010H} is an alias.  The Keplerian orbital elements are reported in Table~\ref{tab:hd156668}, along with the predicted transit window.  The eccentricity was held to zero, as expected from tidal dissipation, following \cite{2010H}.  \cite{2010H} ``filtered" the data by simultaneously fitting a two-planet model and a linear trend. They state that the ''second planet" is a form of high-pass filter, not necessarily an actual planet. We do not fit a linear trend or additional planets in our reported fit and do not subtract them out in Fig.~\ref{fig:hd156668}. However, we have confirmed that our results hold if we do.

\begin{deluxetable*}{ l | l l l l l l l l l l l l l l l l l l }
\tabletypesize{\tiny}
\tablecolumns{19}
\tablewidth{\textwidth}
\tablecaption{\label{tab:hd156668}New parameters for HD~156668,$e_{\rm b}=0$.\tablenotemark{a}}
\tablehead{
&\colhead{$K$}&\colhead{$M\sin i$}&\colhead{$P$}&\colhead{$a$}&\colhead{$e$}&\colhead{$\omega$}&\colhead{$\lambda$}&\colhead{$V$}
\\
&\colhead{ms$^{-1}$}&\colhead{M$_{Earth}$}&\colhead{days}&\colhead{AU}&\colhead{}&\colhead{deg}&\colhead{deg}&\colhead{ms$^{-1}$}
}
\startdata
b&$2.2(3)$&$3.1(4)$&$1.26984(7)$&$0.0211(2)$&$0.000(0)$&$0.(0)$&$136.(19)$\\
&&&&&&&&$-0.4(2)$
\enddata
\tablenotetext{a}{The following gravitational constants were used: $GM_\odot =    0.0002959122082856$, ratio of the sun to Earth $=       332945.51$. The mass of the star was assumed to be 0.77 solar masses. Formal errors from the Levenberg-Marquardt algorithm are given in parentheses, referring to the final digit(s). }
\tablecomments{Data are the Keck data presented by \cite{2010H}.  $T_{\rm epoch}$ is set to the first data point (JD~$2453478.97768$).  These parameters predict a transit epoch of $T_{\rm tr} [\rm{JD}] = 2453478.82(7) + E \times 1.26984(7)$.}
\end{deluxetable*}

\subsection{55 Cnc}

With five discovered planets \citep{2008F}, more than any other extrasolar planetary system, 55 Cnc is a rich environment for study. The first planet was discovered by \citet{1997B}: this planet b has an orbital period of 14.65 days. Five more years of observations revealed two additional planets \citep{2002M}: planet c, with orbital period 44 days, and planet d, with orbital period 5000 days. Measurements from the Hobby-Eberly Telescope \citep{2004M} (HET) revealed, on their own and combined with the Lick measurements by \citet{2002M} and ELODIE measurements by \citet{2004N}, the presence of planet e, with a reported orbital period of  2.8 days. In 2005, in a poster presentation \citep{2005W} and an informally circulated paper \footnote{Available electronically at\\ http://groups.csail.mit.edu/mac/users/wisdom/planet.ps}, Wisdom (hereby referred to as W05) reanalyzed the combined HET, Lick, and ELODIE measurements, found evidence for a 260 day period planet, and questioned whether the reported 2.8 day signal might be an alias of planet c. Finally, \citet{2008F} confirmed the 2.8 day planet e and reported a 260 day planet f based on a decade of  Lick and Keck measurements. They also noted a peak at 460 days and considered whether this peak was an alias of the 260 day planet. 

Because the literature has considered whether they might be aliases and because their periods are in the range where aliases can be the most confusing, planet e and planet f warrant additional consideration.  We confirmed by our analysis that the period of f is correct.  In the following subsection, we apply our method to planet e and find that the 2.8 day period is actually an alias, not of planet c but of a true period of 0.74 days: planet e still exists but its period is actually 0.7 days, not 2.8 days.

\subsubsection{A New Period for 55~Cnc~e}

\begin{figure}[htbp]
   \centering
   \includegraphics[scale=0.9]{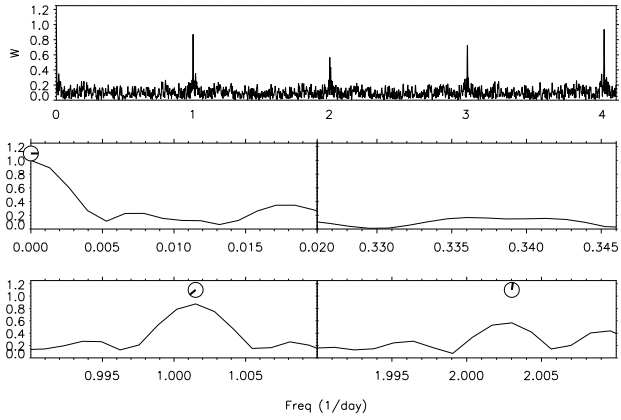}
   \caption{Spectral window function of 55 Cnc for HET data set \citep{2004M}. These features, convolved with a planet's orbital frequency, cause the aliases evident in the periodogram in Fig. \ref{fig:55cnchet}.
   \vspace{0.1 in}}
   \label{fig:win55cnchet}
\end{figure}

\begin{figure}[htbp]
   \centering
      \includegraphics[scale=0.9]{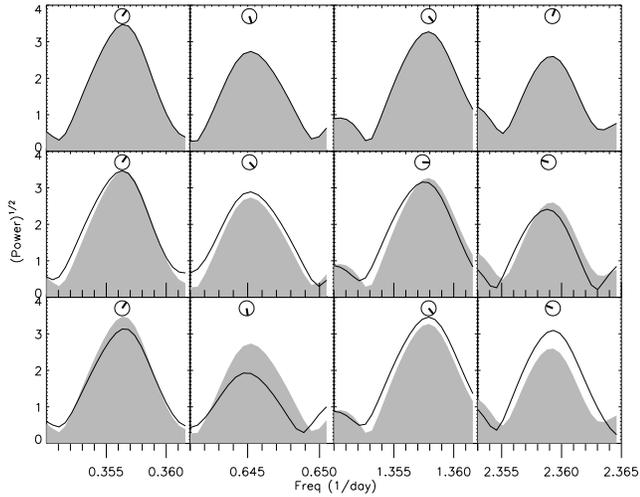}
   \caption{Periodograms of 55 Cnc for planet $e$ only, using only the data from HET \citep{2004M}.  The top row is the periodogram of the data themselves.  The other rows show the periodograms of sinusoids sampled at the times of the real data sets as solid lines; they also repeat the periodogram of the data as a gray background, for comparison.  Dials above the peaks show the phase at each peak. The second row has a sinusoid of the reported frequency.  The third row has a sinusoid of the new frequency. In Rows 2 and 3, each peak results from the convolution of the sinusoidal frequency with the features in the spectral window function in Fig. \ref{fig:win55cnchet}. In this data set, due to noise, neither noiseless candidate frequency matches the data. Note the large phase discrepancies between the reported frequency and the data. Based on this data set alone, the planet's orbital period cannot be unambiguously determined.
   \vspace{0.1 in}}
   \label{fig:55cnchet}
\end{figure}

First, let us look at the discovery data for 55~Cnc~e.  We plot the window function for the data collected by \cite{2004M} using HET in Fig.~\ref{fig:win55cnchet}.  The data spans only 190 days and therefore contains no yearly gaps.  Therefore, no peak in the window function occurs at yr$^{-1}$, and there is no splitting of the daily alias into solar and sidereal days.  We also note that this daily alias has quite a strong value of $\sim 0.8$.  The consequence of that can be seen in Fig.~\ref{fig:55cnchet}, the periodogram using only the HET data.  The top panels are the periodogram of the data themselves.  The peaks at 0.356~day$^{-1}$ and 1.358~day$^{-1}$ are of similar size.  In the middle panels, we sample a noiseless sinusoid with a period, amplitude, and phase matching that of the peak at 0.356~day$^{-1}$.  An alias results at 1.358~day$^{-1}$ at approximately the right height and phase, so \cite{2004M} may have dismissed the latter as an alias, although they did not mention it explicitly.  However, reversing the argument, if we had a noiseless sinusoid with the period, amplitude, and phase of the peak at 1.358~day$^{-1}$ (bottom panels), then its alias nearly matches the peak at 0.356~day$^{-1}$, within the noise.  This is to say, the data of \cite{2004M} cannot distinguish between the two possible periods.

\begin{figure}[htbp]
   \centering
   \includegraphics[scale=0.9]{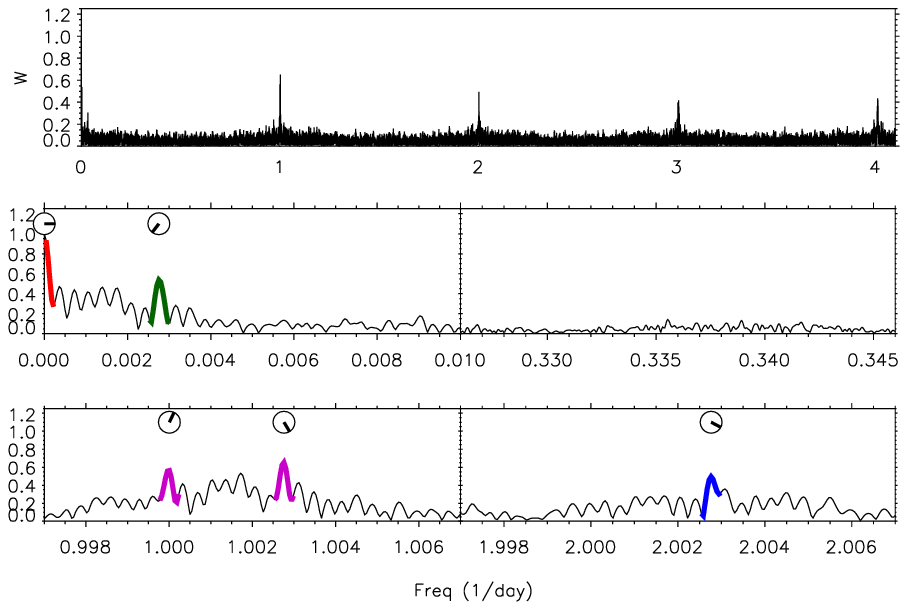}
   \caption{Spectral window function of 55 Cnc for HET data set combined with ELODIE \citep{2004N} and Lick \citep{2002M}. Major features of the spectral window function are colored: red (at 0 day$^{-1}$), green (yearly feature), fuschia (daily features), and blue (two day$^{-1}$). The corresponding aliases these features cause for several candidate frequencies are indicated by these colors in \ref{fig:55cncall}.
   \vspace{0.1 in}
   \label{fig:win55cnccomb}}
\end{figure}

\begin{figure}[htbp]
   \centering
    \includegraphics[scale=0.9]{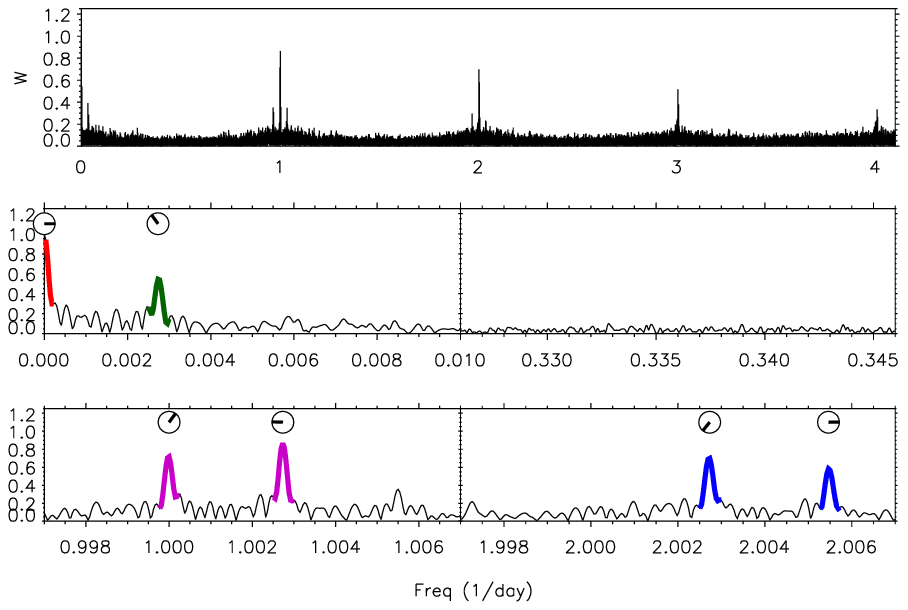}
   \caption{Spectral Window Functions of 55 Cnc for combined Lick and Keck data set \citep{2008F}. Major features of the spectral window function are colored: red (at 0 day$^{-1}$), green (yearly feature), fuschia (daily features), and blue (two day$^{-1}$). The corresponding aliases these features cause for several candidate frequencies are indicated by these colors in \ref{fig:55cncfischer}.
   \vspace{0.1 in}
   \label{fig:win55cncfischer}}
\end{figure}

W05 presented two arguments for why the 2.8 day signal might be an alias. First he noticed that the 2.8 day period is linked to the 44 day period of planet c by a period of 3 days ($\frac{1}{2.8} \approx \frac{1}{3} + \frac{1}{44}$), but noted that there is no reason we would expect an alias to be caused by a 3 day period. In Fig.~\ref{fig:win55cnchet}, \ref{fig:win55cnccomb}, and \ref{fig:win55cncfischer} we demonstrate that there is no peak in the spectral window function at $\frac{1}{3}$ day$^{-1}$ for any of the data sets. Therefore, the 2.8 day signal cannot be an alias of the 44 day signal.

Second, W05 noticed that in the HET data, one peak occurs at 2.808 day, while in the combined data set a pair of peaks occurs at 2.7957 days and 2.8175 day, a splitting of 1 year. In fact, this is just the doublet structure described in section~\ref{sec:daily}.  The combined set spans multiple years, which creates the yr$^{-1}$ spacing in the doublet structure of the daily alias, as shown in Fig.~\ref{fig:win55cnccomb}. Therefore we would actually expect to see this doublet structure in the combined data set but only a single peak at the daily aliases in the HET data set.

So Wisdom was right to suspect that the 2.8 day signal is an alias. It is not an alias of the 44 day planet c but of a planet with true period 0.7 days; the alias is a daily alias (1/2.8 days = 1/0.74 days - 1/days).

\begin{figure}[htbp]
   \centering
   \includegraphics[scale=0.9]{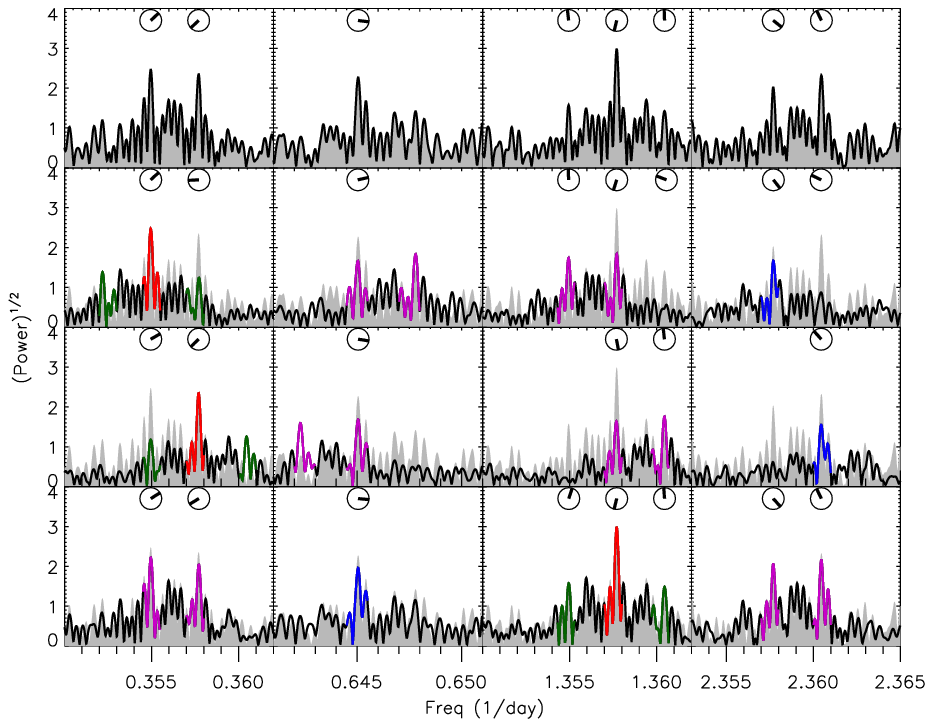}
   \caption{Periodogram of 55 Cnc for planet $e$ only. Dials above the peaks denote their phase. Colors correspond to the feature in the window function that creates the particular alias (see Fig.~\ref{fig:win55cnccomb}), with red being the candidate frequency, the green sidebands yearly aliases, and the fuschia and blue peaks daily and two day$^{-1}$ aliases respectively.  The top row shows the data (HET+ELODIE+Lick). In Row 2  the solid lines show, for these time samplings, the periodogram of a sinusoid of frequency 0.3550 day$^{-1}$.  For Row 3, it is for 0.3577 day$^{-1}$.  For Row 4, it is for 1.3577 day$^{-1}$, our now-favored value. The three candidate frequencies have different types of aliases at different locations, allowing us to break the degeneracy.
   \vspace{0.1 in}
   \label{fig:55cncall}}
\end{figure}

\begin{figure}[htbp]
   \centering
   \includegraphics[scale=0.9]{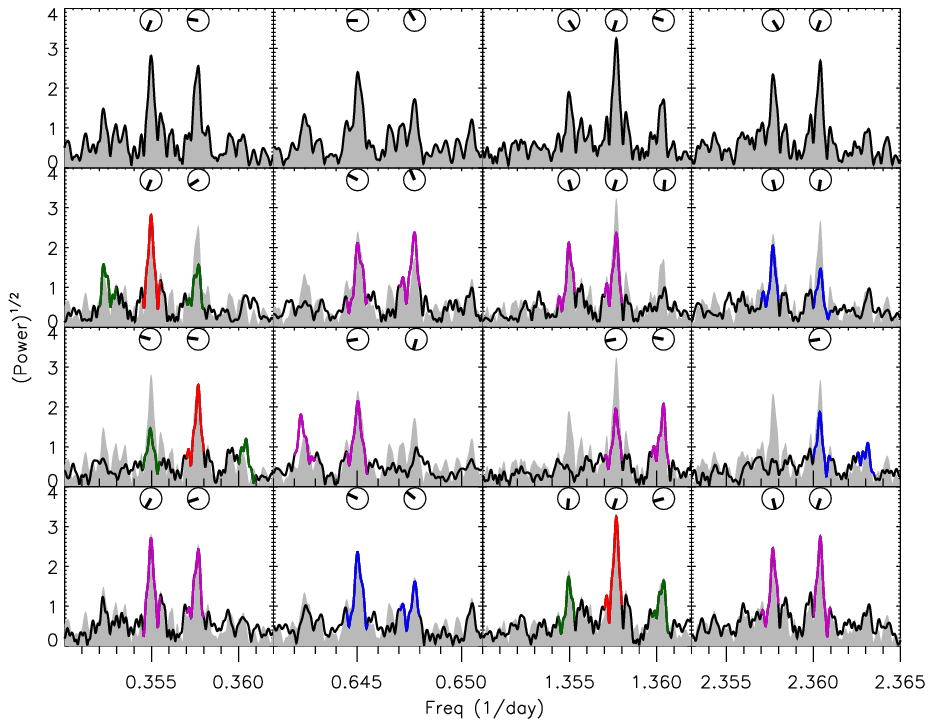}
   \caption{Periodogram of 55 Cnc for planet $e$ only. Dials above the peaks denote their phase. Colors correspond to the feature in the window function that creates the particular alias (see Fig.~\ref{fig:win55cncfischer}), with red being the candidate frequency, the green sidebands yearly aliases, and the fuschia and blue peaks daily and two day$^{-1}$ aliases respectively. Row 1 shows the data (Lick+Keck).  In Row 2  the solid lines show, for these time samplings, the periodogram of a sinusoid of frequency 0.3550 day$^{-1}$.  For Row 3, it is for 0.3577 day$^{-1}$.  For Row 4, it is for 1.3577 day$^{-1}$, our now-favored value. The three candidate frequencies have different types of aliases at different locations, allowing us to break the degeneracy.
   \vspace{0.1 in}
   \label{fig:55cncfischer}}
\end{figure}

With the combined data set, and with new data that has come out with higher precision from Lick and Keck \citep{2008F}, we can confirm with high confidence that the $0.74$~day period is the correct one.  The window functions of these datasets are shown in Fig.~\ref{fig:win55cnccomb} and \ref{fig:win55cncfischer}.  In Fig.~\ref{fig:55cncall} and \ref{fig:55cncfischer} we show the resulting periodograms, after subtracting the signal of planets b, c, and d with a best-fitting Keplerian model.  In both datasets, the true peak at 1.358~day$^{-1}$ is very much higher and the other peaks at various frequencies are fully consistent with being an alias of it.  For instance, in both datasets, doublet structure at the reported frequency shows that it is actually a daily alias.  These peaks are identified for various candidate periods in Tables~\ref{tab:55cnccombplots} and \ref{tab:55cnccombplots2} for the combined data set and ~\ref{tab:55cncfischerplots} and \ref{tab:55cncfischerplots2} for the \citet{2008F} data set. We also performed the same analysis on the combined data set of all four instruments and obtained consistent results. The results are also unambiguous when only the Keck data are used.

With this new period for planet e, we fit a 5-planet Keplerian model to the Keck and Lick data of \citet{2008F}, via the Levenberg-Marquardt algorithm implemented in IDL by \citet{2009Markwardt}.  Following \citet{2008F}, jitter values of 1.5 m/s and 3.0 m/s were adopted for Keck and Lick data, respectively, such that the errors became $\sigma_i^2 = \sigma_{\rm quoted, i}^2 + \sigma_{\rm jitter, i}^2$.  The resulting model fits the data much better than previous results, with the same number of free parameters.  Compare Table \ref{tab:55cncold} and Table \ref{tab:55cnc_newepoch}. The rms is reduced from 6.45 ms$^{-1}$ to 5.91 ms$^{-1}$ (10$\%$) and the $ (\chi_\nu^2)^{1/2} $  is reduced from 1.666 to 1.411 (15$\%$).  We conclude that we have determined the correct period of 55~Cnc~e. 

We use an epoch chosen as the weighted average of the observation times.  The weighting was $1/\sigma_i^2$; this weighting minimizes the correlation between the parameters $P$ and $\lambda$ for each planet. We have confirmed that the rms and $ (\chi_\nu^2)^{1/2} $ we achieve using a weighted epoch, as opposed to using the first data point as the epoch, is identical in the Keplerian case.

With such a small period, we would expect planet e to circularize via tidal dissipation.  Of course, in the presence of perturbations of the other planets, this expectation will not be completely fulfilled.  Nevertheless, we also repeated the fit with the eccentricity of planet e fixed at zero (Table~\ref{tab:55cnc_newepoch_ee0}).

Fitting a self-consistent Newtonian 5-planet model, \citet{2008F} obtained a $ (\chi_\nu^2 )^{1/2}$ of 2.012 and rms of 7.712 ms$^{-1}$, significantly worse than their best Keplerian five-planet model. We performed our own self-consistent Newtonian 5-planet fit using the modified Wisdom-Holman symplectic integrator \citep{1991WH} in SWIFT \citep{1994LD}. Using our newly defined epoch, we obtain $(\chi_\nu^2)^{1/2} $ for both candidate periods of planet e that are statistically indistinguishable from their Keplerian equivalents (Table~\ref{tab:55cncnewtold} and Table~\ref{tab:55cncnewt}). We speculate that the new epoch starts the Levenberg-Marquardt fit closer to the global minimum and strongly recommend choosing the epoch as the weighted average of the observation times, as we have done, instead of the first observation. We have only begun to explore the dynamics of this system and future work adjusting the line of sight inclination of the system and relative inclinations of the planets may result in improved fits and better characterization of the dynamics of this system.

\clearpage

\clearpage

\begin{deluxetable*}{l | l l l l l l l l l }
\tabletypesize{\tiny}
\tablecolumns{         9}
\tablewidth{\textwidth}
\tablecaption{\label{tab:55cnccombplots} 55 Cnc Combined Data Set: expectations from the window function.     \vspace{0.1 in}}
\tablehead{\colhead{Candidate}                                   
&\multicolumn{8}{c}{Window Function Feature}   
\\
\colhead{frequency, $f$}                                   
&\colhead{0.0028-f}
&\colhead{0.0028+f}
&\colhead{1.0000-f}    
&\colhead{1.0027-f}  
&\colhead{1.0000+f}
&\colhead{1.0027+f} 
&\colhead{2.0028-f} 
&\colhead{2.0028+f}      
}
\startdata
\textbf{0.3550}& 0.3522	&0.3578			& 	0.6450	&0.6477	&\textbf{1.3550}&1.3577			&--&-- \\
\textbf{0.3577}&0.3549		&\textbf{0.3605	}	&0.6423	& 0.6450 &1.3577			&\textbf{1.3604}&--&2.3604\\
\textbf{1.3577}&\textbf{1.3549}	& \textbf{1.3605}  &\textbf{0.3550}		          &\textbf{0.3577}&\textbf{2.3577}       &\textbf{2.3604}                     &     --    &--\\
\enddata
\tablecomments{Along the top row are peaks in the window function at frequencies $f_s$ (Fig.~\ref{fig:win55cnccomb}). Each row refers to a candidate frequency $f$; rows 1-3 in this table match to rows 2-4 in Fig.~\ref{fig:55cncall}, respectively.  The cells are frequency values $|f \pm f_s |$ expected for peaks in the periodogram.  If the predicted alias is consistent with a peak in the data in both amplitude and phase, the cell is bolded.  A non-emphasized cell indicates a large discrepancy in amplitude or phase.  For dashed cells, no comparison was done. Units are day$^{-1}$. The frequency of $f=1.3577$~day$^{-1}$ is overwhelmingly the best match to the data.   \vspace{0.1 in}}
\end{deluxetable*}

\begin{deluxetable*}{l | l l l l l l l l l l }
\tabletypesize{\tiny}
\tablecolumns{         11}
\tablewidth{\textwidth}
\tablecaption{\label{tab:55cncfischerplots}55 Cnc \citet{2008F} Data Set: expectations from the window function.   \vspace{0.1 in}}
\tablehead{\colhead{Candidate}                                   
&\multicolumn{10}{c}{Window Function Feature}   
\\
           \colhead{frequency, $f$}                                   
&\colhead{0.0028-f}
&\colhead{0.0028+f}
&\colhead{1.0000-f}    
&\colhead{1.0027-f}  
&\colhead{1.0000+f}
&\colhead{1.0027+f} 
&\colhead{2.0027-f} 
&\colhead{2.0055-f} 
&\colhead{2.0027+f}  
&\colhead{2.0055+f}      
}
\startdata
\textbf{0.3550}& \textbf{0.3522}	&0.3578			& \textbf{0.6450}		&0.6477	& \textbf{1.3550}	& 1.3577			&--&--& \textbf{2.3577}	&2.3605\\
\textbf{0.3577}&0.3549		&\textbf{0.3605	}	&\textbf{0.6423	}	&\textbf{0.6450}	&1.3577			& \textbf{1.3604}	&--& --	&2.3604&\textbf{2.3632}\\
\textbf{1.3577}& \textbf{1.3542}		& \textbf{1.3605}	&\textbf{0.3550}		          &\textbf{0.3577}& \textbf{2.3577}	      &\textbf{2.3604}                     & \textbf{0.	6450}        &\textbf{0.6478}&--&--\\
\enddata
\tablecomments{The format is the same as Table~\ref{tab:55cnccombplots}. Features in the window function are from Fig.~\ref{fig:win55cncfischer}.  The candidate frequencies in rows 1-3 in this table match to rows 2-4 in Fig.~\ref{fig:55cncfischer}, respectively.   \vspace{0.1 in}}
\end{deluxetable*}

\begin{deluxetable*}{l | llllll}
\tabletypesize{\tiny}
\tablecolumns{         7}
\tablewidth{\textwidth}
\tablecaption{\label{tab:55cnccombplots2}55 Cnc Combined Data Set: features in the data periodogram.   \vspace{0.1 in}}
\tablehead{\colhead{Candidate}                                   
&\multicolumn{6}{c}{Major Data Feature}   
\\
           \colhead{frequency, $f$}                        
&\colhead{0.3550}
&\colhead{0.3577}
&\colhead{0.6450}
&\colhead{1.3577}
&\colhead{2.3577}
&\colhead{2.3604}
  }
\startdata
0.3550& \textbf{f} 		&f + 0.0028		& \textbf{1.0000-f}		&\textbf{1.0027+f} 	&\textbf{2.0027+f }	&	\\
0.3577&f-0.0028			& \textbf{f}		&  \textbf{1.0027-f}		&1.0000+f	&		& 2.0027+f	\\ 
1.3577& \textbf{f-1.0027}	& \textbf{f -1 .0000}	& \textbf{2.0027-f} 		& \textbf{f}	&\textbf{1.0000+f} 	& \textbf{1.0027+f}	\\
\enddata
\tablecomments{The top row indicates a major peak seen in the data \emph{near the frequencies where aliases are predicted}. Each row refers to a candidate frequency; rows 1-3 in this table match to rows 2-4 in Fig.~\ref{fig:win55cnccomb}, respectively. If, based on examining the plots, the frequency creates an alias that matches that peak in the data in both amplitude and phase, the cell is bolded. A non-emphasized cell indicates a large discrepancy in amplitude or phase. A blank cell indicates that the candidate frequency does not cause an alias at that frequency. Units are day$^{-1}$. This table shows that a frequency of 1.3577 day$^{-1}$ is best able to account for the peaks in the data.   \vspace{0.1 in}}
\end{deluxetable*}

\begin{deluxetable*}{l | lllllll}
\tabletypesize{\tiny}
\tablecolumns{         8}
\tablewidth{\textwidth}
\tablecaption{\label{tab:55cncfischerplots2}55 Cnc \citet{2008F} Data Set: features in the data periodogram. }
\tablehead{\colhead{Candidate}                                   
&\multicolumn{7}{c}{Major Data Feature}   
\\
           \colhead{frequency, $f$}                        
&\colhead{0.3550}
&\colhead{0.3577}
&\colhead{0.6450}
&\colhead{0.6478}
&\colhead{1.3577}
&\colhead{2.3577}
&\colhead{2.3604}
  }
\startdata
0.3550& \textbf{f} 		&f + 0.0028		& \textbf{1.0000-f}		& 1.0027-f 		&\textbf{1.0027+f} 	&\textbf{2.0027+f }	&2.0054+f 	\\
0.3577&f-0.0028		& \textbf{f}			&  \textbf{1.0027-f}	&				&1.0000+f		&			& 2.0027+f	\\ 
1.3577& \textbf{f-1.0027}	& \textbf{f -1 .0000}	& \textbf{2.0027-f} 		& \textbf{2.0054-f}	& \textbf{f}		&\textbf{1.0000+f} 	& \textbf{1.0027+f}	\\
\enddata
\tablecomments{The format is the same as Table~\ref{tab:55cnccombplots2}. Candidate frequencies in rows 1-3 in this table match to rows 5-8 in Fig.~\ref{fig:win55cncfischer}, respectively.}
\end{deluxetable*}

\begin{deluxetable*}{ l | l l l l l l l l l l l l l l l l l l }
\tabletypesize{\tiny}
\tablecolumns{19}
\tablewidth{\textwidth}
\tablecaption{\label{tab:55cncold} 55 Cnc Keplerian radial velocity fit, $P_e = 2.8$~days.\tablenotemark{a}}
\tablehead{&\colhead{$K$}&\colhead{$M\sin i$}&\colhead{$P$}&\colhead{$a$}&\colhead{$e$}&\colhead{$\omega$}&\colhead{$\lambda$}&\colhead{$V_L$}&\colhead{$V_K$}&\colhead{$\chi^2$}&\colhead{N}&\colhead{$(\chi_\nu^2)^{1/2}$}&\colhead{rms}
\\
&\colhead{ms$^{-1}$}&\colhead{M$_{Jup}$}&\colhead{days}&\colhead{AU}&\colhead{}&\colhead{deg}&\colhead{deg}&\colhead{ms$^{-1}$}&\colhead{ms$^{-1}$}&\colhead{}&\colhead{}&\colhead{}&\colhead{ms$^{-1}$}
}
\startdata
e&$5.2(2)$&$0.0346(16)$&$2.81705(5)$&$0.0382(3)$&$0.066(48)$&$238.(41)$&$86(14)$&\\
b&$71.3(3)$&$0.824(3)$&$14.65164(11)$&$0.1148(8)$&$0.014(4)$&$135.(15)$&$327.4(10)$&\\
c&$10.0(2)$&$0.167(4)$&$44.349(7)$&$0.2402(17)$&$0.09(3)$&$66.(17)$&$312(7)$&\\
f&$5.3(3)$&$0.148(9)$&$259.7(5)$&$0.780(6)$&$0.40(5)$&$182.(9)$&$308(14)$&\\
d&$46.9(4)$&$3.84(4)$&$5191.(53)$&$5.76(6)$&$0.015(9)$&$223.(33)$&$201(4)$&\\
&&&&&&&&$6.8(6)$&$5.9(7)$&$813.2$&$27$&$1.666$&$6.45$
\enddata
\tablenotetext{a}{The following gravitational constants were used: $GM_\odot =    0.0002959122082856$, ratio of the sun to Jupiter $=      1047.35$. The mass of the star was assumed to be 0.94 solar masses. Formal errors from the Levenberg-Marquardt algorithm are given in parentheses, referring to the final digit(s). Masses and semi-major axes are in Jacobian coordinates, as recommended by \citet{2003L}.}
\tablecomments{Data are the Lick and Keck data presented by \cite{2008F}.  $T_{\rm epoch}$ is set to the weighted mean of the observation times (JD~$2453094.762$), which should minimize the correlation in the errors between $P$ and $\lambda$ for each planet. }
\end{deluxetable*}

\begin{deluxetable*}{ l | l l l l l l l l l l l l l l l l l l }
\tabletypesize{\tiny}
\tablecolumns{19}
\tablewidth{\textwidth}
\tablecaption{\label{tab:55cnc_newepoch}55 Cnc Keplerian radial velocity fit, $P_e = 0.74$~days.\tablenotemark{a}}
\tablehead{
&\colhead{$K$}&\colhead{$M\sin i$}&\colhead{$P$}&\colhead{$a$}&\colhead{$e$}&\colhead{$\omega$}&\colhead{$\lambda$}&\colhead{$V_L$}&\colhead{$V_K$}&\colhead{$\chi^2$}&\colhead{N}&\colhead{$(\chi_\nu^2)^{1/2}$}&\colhead{rms}
\\
&\colhead{ms$^{-1}$}&\colhead{M$_{Jup}$}&\colhead{days}&\colhead{AU}&\colhead{}&\colhead{deg}&\colhead{deg}&\colhead{ms$^{-1}$}&\colhead{ms$^{-1}$}&\colhead{}&\colhead{}&\colhead{}&\colhead{ms$^{-1}$}
}
\startdata
e&$6.2(2)$&$0.0261(10)$&$0.736539(3)$&$0.01564(11)$&$0.17(4)$&$177.(13)$&$126(2)$\\
b&$71.4(3)$&$0.826(3)$&$14.65160(11)$&$0.1148(8)$&$0.014(4)$&$146.(15)$&$139.7(2)$\\
c&$10.2(2)$&$0.171(4)$&$44.342(7)$&$0.2402(17)$&$0.05(3)$&$95.(28)$&$90.(2)$\\
f&$5.1(3)$&$0.150(8)$&$259.8(5)$&$0.781(6)$&$0.25(6)$&$180.(12)$&$36(4)$\\
d&$46.6(4)$&$3.83(4)$&$5205.(54)$&$5.77(6)$&$0.024(10)$&$192.(16)$&$222.7(8)$\\
&&&&&&&&$6.7(5)$&$6.5(6)$&$583.1$&$27$&$1.411$&$5.91$
\enddata
\tablecomments{Data are the Lick and Keck data presented by \cite{2008F}.  $T_{\rm epoch}$ is set to the weighted mean of the observation times (JD~$2453094.762$), which should minimize the correlation in the errors between $P$ and $\lambda$ for each planet.  For planet e, these parameters predict a transit epoch of $T_{\rm tr} [\rm{JD}] = 2453094.728(10) + E \times 0.736539(3)$.}
\end{deluxetable*}

\begin{deluxetable*}{ l | l l l l l l l l l l l l l l l l l l }
\tabletypesize{\tiny}
\tablecolumns{19}
\tablewidth{\textwidth}
\tablecaption{\label{tab:55cnc_newepoch_ee0}55 Cnc Keplerian radial velocity fit, $P_{\rm e} = 0.74$~days,$e_{\rm e}=0$.\tablenotemark{a}}
\tablehead{
&\colhead{$K$}&\colhead{$M\sin i$}&\colhead{$P$}&\colhead{$a$}&\colhead{$e$}&\colhead{$\omega$}&\colhead{$\lambda$}&\colhead{$V_L$}&\colhead{$V_K$}&\colhead{$\chi^2$}&\colhead{N}&\colhead{$(\chi_\nu^2)^{1/2}$}&\colhead{rms}
\\
&\colhead{ms$^{-1}$}&\colhead{M$_{Jup}$}&\colhead{days}&\colhead{AU}&\colhead{}&\colhead{deg}&\colhead{deg}&\colhead{ms$^{-1}$}&\colhead{ms$^{-1}$}&\colhead{}&\colhead{}&\colhead{}&\colhead{ms$^{-1}$}
}
\startdata
e&$6.1(2)$&$0.0258(10)$&$0.736540(3)$&$0.01564(11)$&$0.000(0)$&$0.(0)$&$126(2)$\\
b&$71.4(3)$&$0.825(3)$&$14.65158(11)$&$0.1148(8)$&$0.012(4)$&$147.(17)$&$139.7(2)$\\
c&$10.3(2)$&$0.172(4)$&$44.341(7)$&$0.2402(17)$&$0.06(3)$&$99.(23)$&$90.5(15)$\\
f&$5.0(3)$&$0.150(8)$&$260.0(5)$&$0.781(6)$&$0.13(6)$&$180.(21)$&$37(3)$\\
d&$46.7(4)$&$3.83(4)$&$5214.(54)$&$5.77(6)$&$0.029(10)$&$189.(14)$&$222.6(8)$\\
&&&&&&&&$6.8(5)$&$6.3(6)$&$598.1$&$27$&$1.429$&$5.89$
\enddata
\tablecomments{Data are the Lick and Keck data presented by \cite{2008F}.  $T_{\rm epoch} [{\rm JD}]=2453094.762$  Because tidal dissipation has most likely nearly circularized planet the orbit or planet e, here $e_{\rm e}$ is held at zero.  For planet e, these parameters predict a transit epoch of $T_{\rm tr} [\rm{JD}] = 2453094.688(4) + E \times 0.736540(3)$. }
\end{deluxetable*}

\clearpage

\begin{deluxetable*}{ l | l l l l l l l l l l l l l l l l l l }
\tabletypesize{\tiny}
\tablecolumns{19}
\tablewidth{\textwidth}
\tablecaption{\label{tab:55cncnewtold} 55 Cnc dynamical radial velocity fit, $P_e = 2.8$~days.\tablenotemark{a}}
\tablehead{&\colhead{$K$}&\colhead{$M\sin i$}&\colhead{$P$}&\colhead{$a$}&\colhead{$e$}&\colhead{$\omega$}&\colhead{$\lambda$}&\colhead{$V_L$}&\colhead{$V_K$}&\colhead{$\chi^2$}&\colhead{N}&\colhead{$(\chi_\nu^2)^{1/2}$}&\colhead{rms}
\\
&\colhead{ms$^{-1}$}&\colhead{M$_{Jup}$}&\colhead{days}&\colhead{AU}&\colhead{}&\colhead{deg}&\colhead{deg}&\colhead{ms$^{-1}$}&\colhead{ms$^{-1}$}&\colhead{}&\colhead{}&\colhead{}&\colhead{ms$^{-1}$}
}
\startdata
e&$5.1(2)$&$0.0339(16)$&$2.81703(17)$&$0.0382(3)$&$0.09(5)$&$178(4)$&$118(4)$&$$\\
b&$71.4(3)$&$0.825(3)$&$14.6507(4)$&$0.1148(8)$&$0.011(3)$&$143(19)$&$139.7(4)$&$$\\
c&$10.1(2)$&$0.169(4)$&$44.375(10)$&$0.2403(17)$&$0.02(2)$&$359.9(3)$&$88(2)$&$$\\
f&$5.8(3)$&$0.158(8))$&$259.8(4)$&$0.781(6)$&$0.42(4)$&$178(3)$&$33.(3)$&$$\\
d&$47.1(6)$&$3.84(4)$&$5165.(43)$&$5.74(4)$&$0.012(6)$&$279(22)$&$224.0(6)$&$$\\
&&&&&&&&$6.3(5)$&$5.9(6)$&$830.1$&$27$&$1.683$&$6.51$
\enddata
\tablecomments{Data are the Lick and Keck data presented by \cite{2008F}.  $T_{\rm epoch}$ is set to the weighted mean of the observation times (JD~$2453094.762$), which should minimize the correlation in the errors between $P$ and $\lambda$ for each planet. Masses and semi-major axes are in Jacobian coordinates, as recommended by \citet{2003L}.}
\end{deluxetable*}

\begin{deluxetable*}{ l | l l l l l l l l l l l l l l l l l l }
\tabletypesize{\tiny}
\tablecolumns{19}
\tablewidth{\textwidth}
\tablecaption{\label{tab:55cncnewt} 55 Cnc dynamical radial velocity fit, $P_e = 0.74$~days.\tablenotemark{a}}
\tablehead{
&\colhead{$K$}&\colhead{$M\sin i$}&\colhead{$P$}&\colhead{$a$}&\colhead{$e$}&\colhead{$\omega$}&\colhead{$\lambda$}&\colhead{$V_L$}&\colhead{$V_K$}&\colhead{$\chi^2$}&\colhead{N}&\colhead{$(\chi_\nu^2)^{1/2}$}&\colhead{rms}
\\
&\colhead{ms$^{-1}$}&\colhead{M$_{Jup}$}&\colhead{days}&\colhead{AU}&\colhead{}&\colhead{deg}&\colhead{deg}&\colhead{ms$^{-1}$}&\colhead{ms$^{-1}$}&\colhead{}&\colhead{}&\colhead{}&\colhead{ms$^{-1}$}
}
\startdata
e&$6.2(2)$&$0.0260(10)$&$0.736537(13)$&$0.01560(11)$&$0.17(4)$&$181(2)$&$125.(6)$&$$\\
b&$71.4(3)$&$0.825(3)$&$14.6507(4)$&$0.1148(8)$&$0.010(3)$&$139(17)$&$139.6(3)$&$$\\
c&$10.2(2)$&$0.171(4)$&$44.364(7)$&$0.2403(17)$&$0.005(3)$&$252.(41)$&$90.(2)$&$$\\
f&$5.4(3)$&$0.155(8)$&$259.8(5)$&$0.781(6)$&$0.30(5)$&$180.(10))$&$35.(3)$&$$\\
d&$46.8(6)$&$3.82(4)$&$5169.(53)$&$5.74(4)$&$0.014(9)$&$186(8)$&$223.2(7)$&$$\\
&&&&&&&&$6.3(5)$&$6.3(6)$&$591.7$&$27$&$1.421$&$5.96$
\enddata
\tablecomments{Data are the Lick and Keck data presented by \cite{2008F}.  $T_{\rm epoch}$ is set to the weighted mean of the observation times (JD~$2453094.762$), which should minimize the correlation in the errors between $P$ and $\lambda$ for each planet.}
\end{deluxetable*}

\section{Discussion}

\subsection{Summary of Approach}

Aliases result from a convolution between a true physical frequency and the spectral window function, which is created by gaps in the data set due to observational constraints. Our method harnesses features in the window function to distinguish aliases from true frequencies. For a given frequency $f$ and window function peak $f_s$, aliases will occur at $|f \pm f_s |$, where $f_s$ is a feature in the window function. In the ranges where we expect major aliases to occur, we compare the phase and amplitude of aliases predicted by a sinusoid of the candidate frequency sampled to the data, with other known planets subtracted off beforehand. We judge whether the ``pattern" of the predicted aliases matches the data: for example, yearly aliases appear as sidebands of the candidate frequency while daily aliases often appear as a doublet caused by the sidereal and solar day. If all the aliases match in amplitude, phase, and pattern, we can be confident that we have found the true orbital period. If there are discrepancies and the aliases of none of the candidate frequencies match the data, we know that noise prevents us from definitively determining the true period and that follow-up observations are necessary. Misunderstandings about aliases have previously led to incorrect identification of planet's orbital periods, a key parameter in defining the planet's properties, as well as the dynamical behavior of the planets in the system. We have corrected common misconceptions, including that aliases always appear near the frequency of peaks in the window function, that any frequency above 1 cycle/day is necessarily an alias, and that aliases will appear if the data are scrambled or if the true frequency is subtracted out. 

\subsection{Summary of Results}

For two systems, we confirmed previous distinctions between alias and true frequency. The period of GJ~876~d is indeed 1.94 days, not 2.05 days. The period of HD~75898~b is indeed 400 days and the periodogram peak at 200 days is indeed an alias, not a second planet or eccentricity harmonic, the alternative explanations proposed by \citet{2007R}.

For two other systems, we determined that the data are too noisy to allow us to definitely distinguish between alias or true frequency. According to our analysis, it remains unclear whether the period of Gl~581~d is 67 days or 83 days; even a period of ~1 day cannot be ruled out. It also remains unclear whether HD~73526 contains two planets with orbital periods 187.5 and 376.9 days, locked in a 2:1 resonance, or whether one of the periods is actually 127 days. Further observations of these systems are required, preferably at times that reduce the aliasing.

For a final pair of systems, we determined the reported orbital period was incorrect, due to mistaking a daily alias for the true frequency. According to our analysis, the orbital period of HD~156668~b is actually 1.2699 days, not 4.6455 days. The orbital period 55~Cnc~e is 0.7365 days, not 2.817 days. The standard, general-purpose software \emph{SigSpec} mentioned in the introduction \citep{2007Reegen,2010R} agrees with our orbital period distinctions (we used the parameters: depth=2, par = 0.2 and par = 0.5, and a frequency upper limit of 2 day${-1}$).

\subsection{Implications for 55~Cnc~e}

What are the implications of an updated period for the innermost planet of 55 Cnc?

First, it dramatically lowers the effective noise when determining the parameters of the planetary system.  \citet{2008F} reported independent Keplerian fits with rms of 6.74 m/s, and a self-consistent dynamical fit with rms of 7.712 m/s. Our Keplerian fit achieves rms of 5.91 m/s, and our self-consistent coplanar dynamical fit achieves rms of 5.96 m/s. By adjusting the inclination of the system relative to our line-of-sight and the planets' mutual inclinations, an even better self-consistent might be possible. Therefore perturbations might be directly detected via a lower rms when interactions among the planets are included, and the architecture of the system further constrained.  We have just begun exploring this avenue.

Second, 55~Cnc~e itself can now be searched for transits at the new period, with high a priori probability of $\sim25$\%.  Given the period and phase of the radial-velocity signal, we report predicted transit epochs in Tables \ref{tab:55cnc_newepoch} and \ref{tab:55cnc_newepoch_ee0}.  The predictions differ because the latter assumes zero eccentricity, and the formally significant value of $e_{\rm e}$ matters.  Nevertheless, folding the systematic uncertainty related to eccentricity into the predicted transit time, we still can predict transit times good to $\sigma_T\simeq1$~hour in 2010.  This search can be accomplished simply by folding the photometric data reported by \cite{2008F} at the new ephemeris.  Gregory Henry (priv. comm.) has made such a search, finds no positive signal, and constrains putative transits in the period range $0.7-0.8$~days to a depth $<0.7$~mmag, or $>2.6$~$R_\oplus$.  Earth-composition models of super-Earths predict a radius $\sim1.9$~$R_\oplus$ \citep{2006V}, so a search at higher precision is certainly worthwhile.

Third, even apart from a transit, this super-Earth must be very hot, as it is very close-in to a solar-type star. Following \citet{2009L}, we find that the substellar point could be up to 2750 K, if the insolation is absorbed then reradiated locally.  We would naively expect that the enormous radiation this planet takes in would evaporate any atmosphere (e.g., \citealt{2010J}).  Moreover, the host star is also very bright as seen from Earth.  Therefore it might be useful to look for its phase curve with Spitzer, to detect or rule out an atmosphere \citep{2009SD}.  Another attractive possibility is probing a magma ocean, which may exist because of the irradiation \citep{2009G,2010G}, but this may require transit measurements.

Fourth, the presence of the other 4 planets surely injects a non-zero eccentricity into this tidally-dissipating planet.  Its expected value remains to be calculated, but will likely be on the order of $10^{-4}$.  This forced eccentricity could stimulate considerable geologic activity --- it might be a ``super-Io'' \citep{2010B}.

\subsection{Observational Strategies for Mitigating Aliases}

Can aliases be prevented or mitigated by the choice of observation times? Constraints on when the star is visible at night necessarily result in gaps in the data that cause aliases. However, we encourage observers to engage in ``window carpentry" \citep{1982S} by observing the star during the greatest span of the sidereal and solar day possible, not just when the star transits the meridian. Unfortunately, observing stars as they rise and set poses a challenge for observers, who minimize slew time\footnote{If the slew time exceeds the read-out time, fewer observations may be gathered per night.  However, the wise spacing of observation times can more than make up for this through disambiguation of alias frequencies using fewer data points.} and thus maximize the number of stars observed per night by observing at the meridian for the majority of the night. This observing strategy (Fig.~\ref{fig:mw1}b) results in strong daily and yearly aliases (ex. Fig.~\ref{fig:wingl581}). Another strategy is to start in the west and gradually move east over the course of the night (Fig.~\ref{fig:mw1}c), observing as much of the sky as possible. This strategy reduces yearly aliases but  sidereal daily aliases remain strong (ex. Fig.~\ref{fig:winhd156668} and  \ref{fig:win55cncfischer}). To reduce sidereal daily aliases, we recommend the following procedure.  Start the telescope somewhere west of the meridian (randomized from night to night) and move east to cover half the sky over the course of half the night. Then make one large slew to the place the telescope started and re-observe the same portion of the sky. Some stars will gain the advantage of being observed twice in one night.  Moreover, when the data are folded at the mean sampling period, they still show some variety in phase of observation, which is needed to reduce window function peaks and de-alias candidate periods. However, another consideration is that at higher air mass, both the extinction is greater and the seeing is worse. The increased atmospheric attenuation means a longer integration time is required, reducing the number of stars that can be observed, while the seeing increases the measurement errors. For a particular set of stars, observers can work out a slew pattern that will maximize the number of stars observed while minimizing aliasing. \citet{2006S} present a clever method for determining the optimal sampling when period searching using satellite telescopes or a longitude-distributed network that can observe continuously. Unfortunately this strategy is impractical to implement using a single telescope on the ground. \citet{2008Ford} presents useful adaptive scheduling algorithms for observing multiple targets that can be parametrized to reduce aliasing.

We suggest taking advantage of any unusual time windows: for example, the rare granted dark time or time at the beginning or end of another observer's night. Observers focusing on a large group of stars can determine which star would most benefit from this unusual time by calculating the window function with the new observation times added or, in the case of a planet with two candidate periods, determining for which system the observation times would best distinguish between two candidate orbits. We also suggest that it would be beneficial to observe stars using telescopes in two or more locations at different latitude and longitudes (ex. Fig.~\ref{fig:win55cnccomb}).
 
At the stage of data analysis, we encourage the use of our method to distinguish true frequencies from aliases, crucial for the correct characterization of the planet. As astronomers push to observing lower mass planets and modeling planets near the noise limit, they cannot assume that the highest peak in the periodogram -- or even the best Keplerian fit -- corresponds to the true orbital period. Only by harnessing features in the window function to compare the amplitude, phase, and pattern of an assortment of predicted aliases to the data can we distinguish the planet's true orbital frequency -- or determine that more observations are needed.

\subsection{Conclusion}

Knowing a planet's correct orbital period is essential for accurately characterizing it. By Kepler's law, the planet's distance from the star increases as its orbital period increases. Therefore the planet's orbital period sets its temperature: too hot, too cold, or just right for life. The planet's inferred mass, as calculated from the radial velocity amplitude, increases as the period decreases -- a closer planet needs less mass to exert a given force on the star -- so a difference in orbital period may be the difference between an Earth analog and a super-Earth. In the case of multi-planet systems, the spacing of the planets determines their mutual interactions: therefore a difference in orbital period may be the difference between a precariously placed planet and one locked deep in a stabilizing resonance. The signal of a planet's eccentricity is contained in the harmonics of the planet's orbital period: therefore a difference in orbital period may be the difference between a planet that formed in situ and a planet violently scattered, a calm planet that has long been tidally circularized or a planet erupting with volcanoes due to tidal dissipation. But periods that correspond to totally different worlds are only subtly distinguishable in the radial velocity signal. Such are the machinations of aliases.

Through our method, astronomers can confirm a planet's orbital period or determine that noise prevents a definitive distinction. In the latter case, follow-up observations taken according to the suggestions above should eventually allow the true period to be determined. Ironically, Earth's own rotational and orbital period make it challenging to uncover the orbital period of other worlds, particularly Earth analogs. But by better understanding of digital signal processing, we can mitigate the deleterious effects of the inevitable sunrise and starset.

\acknowledgements We gratefully acknowledge support from the Harvard University Department of Astronomy and from the Michelson Fellowship, supported by the National Aeronautics and Space Administration and administered by the NASA Exoplanet Science Center. We thank Debra Fischer, Andrew Howard, Greg Henry, Barbara McArthur, and Geoff Marcy for helpful discussions. We thank two anonymous referees and editor Steven Kawaler for their insightful feedback. We are also grateful for helpful comments from Eric Agol, Christopher Burke, and Scott Tremaine.

\bibliography{aliases} \bibliographystyle{apj} 

\end{document}